\documentclass[aip,amsmath,amssymb,reprint, onecolumn]{revtex4-1}

\usepackage[utf8]{inputenc}
\usepackage[T1]{fontenc}
\usepackage{mathptmx}

\usepackage{times,amsmath}
\usepackage{epsfig}
\usepackage{color}
\usepackage{longtable}
\usepackage{ulem}
\usepackage{graphicx}% Include figure files
\usepackage{dcolumn}% Align table columns on decimal point
\usepackage{bm}% bold math
\usepackage{bookmark}
\usepackage{tabularx}% bold math
\usepackage{hyperref}
\usepackage{multirow}
\hypersetup{colorlinks=true, citecolor=blue, filecolor=blue, linkcolor=blue, urlcolor=blue}

\begin{document}
\title{Neural Networks Potential from the Bispectrum Component: A Case Study on Crystalline Silicon}

\author{Howard Yanxon}
\affiliation{Department of Physics and Astronomy, University of Nevada, Las Vegas, NV 89154, USA}
\affiliation{Materials Science Division, Lawrence Livermore National Laboratory, Livermore, California 94550, USA}

\author{David Zagaceta}
\affiliation{Department of Physics and Astronomy, University of Nevada, Las Vegas, NV 89154, USA}

\author{Brandon C. Wood}
\affiliation{Materials Science Division, Lawrence Livermore National Laboratory, Livermore, California 94550, USA}

\author{Qiang Zhu}
\email{qiang.zhu@unlv.edu}
\affiliation{Department of Physics and Astronomy, University of Nevada, Las Vegas, NV 89154, USA}

\date{\today}
\begin{abstract}
In this article, we present a systematic study in developing machine learning force fields (MLFF) for crystalline silicon. While the main-stream approach of fitting a MLFF is to use a small and localized training sets from molecular dynamics simulation, it is unlikely to cover the global feature of the potential energy surface. To remedy this issue, we used randomly generated symmetrical crystal structures to train a more general Si-MLFF. Further, we performed substantial benchmarks among different choices of materials descriptors and regression techniques on two different sets of silicon data. Our results show that neural network potential fitting with bispectrum coefficients as the descriptor is a feasible method for obtaining accurate and transferable MLFF.
\end{abstract}
\maketitle

\section{INTRODUCTION}
Atomistic modeling methods such as molecular dynamics (MD) or Monte Carlo (MC) play important roles in investigating time-dependent physical and chemical processes. In these methods, energy and forces need to be recalculated iteratively as the atomic configuration evolves. Consequently, atomistic simulations crucially depend on the accuracy of the underlying potential energy surface (PES). Modern quantum mechanical modeling based on density functional theory (DFT) can consistently generate accurate energetic descriptions for many solid systems \cite{lejaeghere2016reproducibility}. However, MD simulations based on DFT suffer from the highly demanding computational cost. The simulations are only suitable to model a system with up to a few thousands of atoms at tens of picoseconds. On the other hand, classical force field (FF) method is widely employed to simulate materials with millions of atoms at hundreds of nanoseconds. This method has enabled many explorations that lead to revealing interesting physical and chemical phenomena \cite{berber2000unusually, yamakov2002dislocation, yamakov2004deformation}. However, the construction of a reliable PES by classical FF method remains problematic. In developing classical FF, a set of parameters are fitted to a few DFT and/or experimental data to compute the potential energy of a system given an analytic functional form. Due to the constraints on the functional form and the limitation of training data set, the accuracy of classical FF is not dependable.

Meanwhile, \textit{in-silico} materials discovery requires an accurate yet efficient energy model to screen materials' properties in high-throughput manner. In the past decade, the discoveries of new materials have been highly driven by advanced structure prediction methods such as crystal structure prediction (CSP) \cite{oganov2019structure} and data mining \cite{Curtarolo-2013}. In both cases, DFT method is used to perform geometry relaxation and energy evaluation. Despite the power of the current supercomputer, the computational cost for DFT simulation remains a bottleneck to many important and fascinating puzzles in materials science. Ideally, an approach that preserves DFT accuracy without sacrificing the computational cost is desirable.

To resolve the limitations described above, many efforts have been devoted towards establishing machine learning force field (MLFF) method. Compared to the DFT method, MLFF approach demands far lower computational cost (2-4 orders of magnitude lower) while retaining accuracy at the DFT level. The power of MLFF method is illustrated by many applications to a range of materials \cite{bartok2013machine, artrith2011high, khaliullin2011nucleation, behler2008metadynamics}. A large amount of DFT data (structures, energy, forces, and stresses) are required to develop an accurate MLFF. The structures must be represented by appropriate descriptors (high-dimensional real valued array) in order to identify the similarities and/or dissimilarities in the atomic environments. In MLFF fitting, a variety of regression techniques are used to correlate between the descriptor and energy/forces. Several machine learning techniques for developing MLFF had been successfully implemented: linear/polynomial regression \cite{thompson2015spectral, wood2018extending, pozdnyakov2019fast, MTP}, Gaussian process regression \cite{bartok2010gaussian, bartok2015g}, and high-dimensional neural network potential (NNP) \cite{behler2007generalized, behler2015constructing}. A benchmark study of these machine learning methods had been carried out for performance and cost inspections to many elemental systems \cite{zuo2019performance}. Nevertheless, many of the published MLFFs lack of transferability/versatility which is crucial in crystal structure prediction.

In the past few years, many researchers have attempted to improve transferability for many different systems \cite{behler2008metadynamics, Hajinazar-PRB-2017, deringer2018data, podryabinkin2019accelerating, Jacobsen-PRL-2018, zeni2018building, Bartok-PRX-2018}. Two approaches, including advanced sampling and structure prediction, have recently become popular. One is to force ordinary MD simulations to escape from the already explored equilibrium states \cite{herr2018metadynamics, huang2018atomic}, while the other attempts to identify the low energy configurations by sampling many different basins mostly based on geometry optimizations. A heterogeneous training data set---diversity in structural types---enhance transferability across different types of structures, curing the extrapolation problem \cite{zeni2018building, Bartok-PRX-2018}. Zeni \textit{et al.} \cite{zeni2018building} achieved a good trade-off between transferability and overall accuracy by applying Gaussian process regression with diverse data set (including high temperature structures). Similarly, many physical properties were reproduced within 10\% relative error to the DFT \cite{Bartok-PRX-2018}. Hajinazar \textit{et al.} employed a structure prediction technique to generate more diverse data sets than the common, less diverse, data set generated with MD-based approach \cite{Hajinazar-PRB-2017}. In addition, it was proposed that the generation of MLFFs could be performed in conjunction with structure prediction processes. The active learning approach in constructing MLFFs on-the-fly was employed automatically to deal with extrapolation outside the training domain. Then, the MLFFs replaced the DFT gradually for structural relaxation s and energy evaluations with much lower computational cost. Active learning technique had been successfully applied to predict PES reconstructions of several challenging elemental systems \cite{deringer2018data, podryabinkin2019accelerating} and multi-component system \cite{Jacobsen-PRL-2018}. For instance, Deringer \textit{et. al.} \cite{deringer2018data} used Gaussian process regression combined with random structure searching (RSS) algorithms to systematically construct an interatomic potential for boron; Podryabinkin \textit{et. al.} \cite{podryabinkin2019accelerating} employed the evolutionary algorithm USPEX to build the machine-learning interatomic potentials for several elemental allotropes; similar ideas were also applied to investigate the surface reconstructions \cite{Jacobsen-PRL-2018} and nano particles \cite{Kolsbjerg-2018-PRB}.

In this report, we will discuss about our attempts in developing accurate and transferable MLFF for elemental silicon as the prototypical system. Many silicon MLFFs had been developed using the training data sets obtained by running MD simulations and selecting known structural prototypes manually \cite{behler2008metadynamics, Bartok-PRB-2013, bartok2010gaussian, Li-PRL-2015, Bartok-PRX-2018, BabaeiSi2019, Bonati-PRL-2018, zuo2019performance}. These configurations from MD trajectories tend to possess strong correlations with the initial geometry. Hence, the resulting MLFFs can only describe a few energy basins of the entire PES. We believe that there are two main factors can influence the transferability of the MLFF. First, training data set generated with high-throughput structure prediction method can enhance the transferability. Here, we generate a diverse silicon data set by using our in-house code, PyXtal \cite{pyxtal}---a Python package for random crystal structure generation. The DFT-quality data set spans a large space in the PES covering many energy basins, and the DFT setting is provided in section \ref{setting}. Second, we enable a machine learning infrastructure that allows Behler-Parrinello descriptors and bispectrum coefficients descriptors to be trained with generalized linear regression and neural network. The details of the descriptors and the regression techniques are available in section \ref{sec:descriptors} and section \ref{machinelearning}, respectively. Finally, we will systematically construct the NNP with bispectrum coefficients, as the descriptors in section \ref{sec:results}.

\section{Computational Methodologies} \label{sec:method}
\subsection{\textit{Ab initio} calculation}
\label{setting}
\textit{Ab initio} calculations are neccessary to provide the training data set for MLFF development. In this study, we employed PyXtal \cite{pyxtal} software to generate several thousands of structural configurations. For each configuration, the total energy and forces were calculated at DFT level through the ASE package \cite{larsen2017atomic}. ASE provides interface to the VASP code \cite{kresse1996efficient} within projector augmented wave methodology \cite{blochl1994projector} to perform geometry relaxations. In our calculation, we used the PBE-GGA \cite{PBE-PRL-1996} as the exchange-correlation functional with an energy cutoff of 600 eV and a $\Gamma$-centered KSPACING of 0.15. 

\subsection{Descriptors}\label{sec:descriptors}
Descriptors, as the unique numerical representations of atomic structures, play an essential part in constructing MLFF. It is crucial for a descriptor to be able to distinguish the local environments of atomic structures. While the most common choice of representation by atomic coordinates is convenient, but it poorly describes the structural environments. The Cartesian coordinates of a crystal structure can change through translational or rotational operation, while the energy remains invariant. Thus, physically meaningful descriptors must be unaffected by these alterations to the structural environment, and any permutation of atoms should not change the descriptors. Additionally, the descriptors must be continuously differentiable within the domain of local atomic environment. In the last decade, the atom-centered descriptors, which probe the atomic environment by its neighboring vectors, become popular because they fit the criteria. The descriptors usually operate within a cutoff function to ensure that the descriptors smoothly vanish to zero at a given cutoff radius, $R_\textrm{c}$. A popular cutoff function choice is the so called cosine cutoff function. The function is expressed in the following:
\begin{equation} \label{cosine}
    f_\textrm{c}(R_{ij}; R_\textrm{c}) = 
        \begin{cases}
        \frac{1}{2}\left[\cos\left(\frac{\pi R_{ij}}{R_\textrm{c}}\right) + 1\right]     & R_{ij} \leq R_\textrm{c}\\
        0                                                                       & \text{otherwise}
        \end{cases}
\end{equation}
where $R_{ij}$ is the distance between the center atom $i$ and the neighbor atom $j$. 

Among the atom-centered descriptors, Behler-Parrinello descriptors \cite{behler2007generalized} and bispectrum coefficients \cite{bartok2010gaussian} are widely used in the materials modelling community. Their definitions will be discussed briefly as follows.

\subsubsection{Behler-Parrinello descriptors}\label{BehPar}
Behler-Parrinello descriptors are used regularly to represent the local atomic environments of crystal structures in NNP development. Commonly used Behler-Parrinello descriptors are two-body ($G^2$) and three-body ($G^4$) symmetry functions:
\begin{equation}\label{G2}
    G^2_i = \sum_{j\neq i} e^{-\eta (R_{ij}-R_\textrm{s})^2} f_\textrm{c}(R_{ij})
\end{equation}
\begin{equation} \label{G4}
\begin{aligned}
    G^4_i = &2^{1-\zeta}\sum_{j\neq i} \sum_{k \neq i, j} (1+\lambda \cos \theta_{ijk})^{\zeta} \cdot \\
    &e^{-\eta (R_{ij}^2 + R_{ik}^2 + R_{jk}^2)} \cdot f_\textrm{c}(R_{ij}) \cdot f_\textrm{c}(R_{ik}) \cdot f_\textrm{c}(R_{jk})
\end{aligned}
\end{equation}
$G^2$ is mainly designed to capture the radial environment while $G^4$ is used for describing the angular part by including the three-body $ijk$ terms. $R_\textrm{s}$ shifts the center of the Gaussian functions to a certain radius resulting in spherical shell with the Gaussian width of $\eta$. $\zeta$ controls the angular resolution, and $\lambda$ usually takes the value of +1 and -1 for inverting the cosine function. The cutoff function ($f_\textrm{c}$) is consistent with Eq. \ref{cosine}. There is a set of $G^2_i$ and $G^4_i$ descriptors specifying the center atom $i$ in relation to the neighboring atoms $j$ in terms of radial and angular parts. For a real material system, this set of parameters need to be optimized by a more extensive search \cite{Gastegger-2018-JCP, Giulio-JCP-2018, Huan-2017-NCM, Gao-JCP-2019}.

\subsubsection{Bispectrum Coefficients}\label{bispectrum}
Similar to Behler-Parrinello descriptors, SO(4) bispectrum can be used to represent the local atomic environments. It was first introduced by Bart\'ok \textit{et al.} for the training of machine learning FF (MLFF) on the elemental systems of Group IVA\cite{bartok2010gaussian}. A detailed study of SO(4) bispectrum as a descriptor along with several alternative implementations (SO(3) bispectrum, angular Fourier series, and SOAP kernel) is available in Ref. \cite{Bartok-PRB-2013}. Later, Thompson \textit{et al.} proposed the spectral neighbor analysis method (SNAP) method and demonstrated that the SO(4) bispectrum could achieve satisfactory accuracy based on the simple linear \cite{thompson2015spectral} and quadratic regressions \cite{wood2018extending}. Following the original work, the expression of SO(4) bispectrum is formed by the expansion coefficients of 4D hyperspherical harmonics:
\begin{equation}
\begin{aligned}
    &B_{i}^{l_1,l_2,l} = \sum_{m, m'=-l}^{l} (c^{l}_{m',m})^* \\
    &\sum_{m_1, m_1'=-l_1}^{l_1} \sum_{m_2, m_2'=-l_2}^{l_2}c^{l_1}_{m_1',m_1} c^{l_2}_{m_2',m_2} H^{l, m, m'}_{l_1,m_1,m_1',l_2,m_2,m_2'}
\end{aligned} 
\end{equation}
where $H^{l_1, l_2, l}_{m_1',m_2',m',m_1,m_2,m}$ is the analog to the Clebsch-Gordan coefficients on the 3-sphere. In application, it is the product of two ordinary Clebsch-Gordan coefficients on the 2-sphere. $c^{l,m}_{l_1, m_1, l_2, m_2}$ are the expansion coefficients from the hyperspherical harmonics ($U^{l}_{m',m}$) functions that are projected from the atomic neighborhood density within a cutoff radius onto the surface of four-dimensional sphere:
\begin{equation}
    \rho = \sum_{l=0}^{+\infty}\sum_{m=-l}^{+l}\sum_{m'=-l}^{+l}c^l_{m',m}U^{l}_{m',m}
\end{equation}
where the expansion coefficients are defined as
\begin{equation}
    c^l_{m',m} = \left<U^l_{m',m}|\rho\right>
\end{equation}

In this work, our implementation of SO(4) bispectrum or bispectrum descriptor is very similar to the SNAP method \cite{thompson2015spectral} that is implemented in the LAMMPS code \cite{LAMMPS}. However, we introduce another method to calculate the hyperspherical harmonics and their gradients\cite{PhysRevD.87.104006}. The benefit of this method is that it allows for the removal of singularities at the north and south poles of the 3-sphere that exist in the traditional implementation. Furthermore, we also include an option to normalize the expansion coefficients from the hyperspherical harmonics, where the normalization factor is $\frac{\sqrt{2l+1}}{4\pi}$. The impacts of normalization on the MLFF training will be discussed later in section \ref{new_strategy}.

\subsection{Machine Learning Force Field Fitting}\label{machinelearning}
The construction of the total energy ($E_\textrm{total}$) of a structure can be obtained by the summation of atomic energy ($E_i$) evaluated from atom-centered descriptors, $\boldsymbol{X}_i$:
\begin{equation}\label{total_E}
    E_\textrm{total} = \sum_i ^{\textrm{all atoms}} E_i(\boldsymbol{X}_i)
\end{equation}
The atomic energy contributions depend on the local structural environment within a cutoff radius with respect to the center atom $i$. Furthermore, accurate representation of PES is also dependent on the contributions of forces. The force acted on atom j can be expressed by the negative gradient of the energy with respect to its atomic positions ($\boldsymbol{r}_j$):
\begin{equation}\label{force}
     \boldsymbol{F}_j=-\sum_i ^{\textrm{all atoms}} \frac{\partial E_i(\boldsymbol{X}_{i})}{\partial \boldsymbol{X}_{i}} \cdot \frac{\partial
    \boldsymbol{X}_{i}}{\partial \boldsymbol{r}_j}
\end{equation}
The functional forms of $E$ and $F$ are fully dependent on the regression algorithm. Generalized linear regression and neural network (NN) regression will be discussed in the following sections.

\subsubsection{Generalized Linear Regression}\label{mdLR}

Linear regression is the most fundamental approach in curve fitting. In this context, each atomic energy is assumed to be linearly correlated with the descriptors. Thus, the total energy can be expressed as follows,
\begin{equation}\label{PolyEnergy}
    E_{\textrm{total}} = \gamma_0 + \boldsymbol{\gamma} \cdot \sum^{N}_{i=1}\boldsymbol{X}_i,
\end{equation}
where $\gamma_0$ and $\boldsymbol{\gamma}$ are the weights presented in scalar and vector forms, and $N$ is the total atoms in a structure.

In general, the total energy can be described as a generalized linear regression with extended polynomial terms. Below is a version to the second-order (quadratic) expansion in the Taylor series:
\begin{equation}\label{PolyEnergy}
    E_{\textrm{total}} = \gamma_0 + \boldsymbol{\gamma} \cdot \sum^{N}_{i=1}\boldsymbol{X}_i + \frac{1}{2}\sum^{N}_{i=1}\boldsymbol{X}_i^T\cdot \boldsymbol{\Gamma} \cdot \boldsymbol{X}_i 
\end{equation}
where $\boldsymbol{A}$ is the symmetric weight matrix (i.e. $\boldsymbol{A}_{12} = \boldsymbol{A}_{21}$) describing the quadratic terms. From linear to quadratic regression, the size of weight coefficients increases from $N+1$ to $(N+1)(N+2)/2$. Indeed, the energy can be further expanded to higher order. However, we restrict it to the second-order expansion due to the drastic increase in the size of weight coefficients.

Correspondingly, the force of an atom $j$ can be expressed in this form by expanding the terms in Eq. \ref{force} with Eq. \ref{PolyEnergy}:
\begin{equation}\label{PolyForce}
    \boldsymbol{F}_j = \sum^{N}_{i=1} \bigg(-\boldsymbol{\gamma} \cdot \frac{\partial \boldsymbol{X}_i}{\partial \boldsymbol{r}_j} - 
    \frac{1}{2} \bigg[\frac{\partial \boldsymbol{X}_i^T}{\partial \boldsymbol{r}_j} \cdot \boldsymbol{\Gamma} \cdot \boldsymbol{X}_i + \boldsymbol{X}_i^T \cdot \boldsymbol{\Gamma} \cdot \frac{\partial \boldsymbol{X}_i}{\partial \boldsymbol{r}_j} \bigg]\bigg)
\end{equation}
Both energy and force terms have a linear correlation with the expanded descriptors through a set of weight coefficients $\{\gamma_0, \boldsymbol{\gamma}_1, ..., \boldsymbol{\gamma}_N, \boldsymbol{\Gamma}_{11}, \boldsymbol{\Gamma}_{12}, ..., \boldsymbol{\Gamma}_{NN}\}$. For convenience, we call the set of coefficients as $\boldsymbol{w}$ from now on. To obtain the best $\boldsymbol{w}$, we solve the objective cost function following the least squares formula for both energy and force,
\begin{equation}\label{loss}
\begin{split}
    \Delta = \frac{1}{2s}\sum_{i=1}^s\Bigg[\bigg(\frac{E_i - E^{\textrm{Ref}}_i}{N_{\textrm{atom}}^i}\bigg)^2 +
             \frac{\beta} {3N_{\textrm{atom}}^i}\sum_{j=1}^{3N_{\textrm{atom}}^i}
    (F_{i, j} - F_{i, j}^{\textrm{Ref}})^2 \Bigg]
\end{split}
\end{equation}
where $s$ is the total number of structures, $i$ loops over all structures, and $j$ loops over all atoms for each structure $i$ in all three directions. $N^{\textrm{atom}}_i$ is the total number of atoms in the $i$-th structure. $\beta$ is the force coefficient. It balances the energy and force contributions due to the number of force components is much larger. The cost function compares the predicted values obtained from the regression ($E_i$ and $F_{i, j}$) to the true values of $E^{\textrm{Ref}}$ and $F_{i, j}^{\textrm{Ref}}$. 

To prevent overfitting, it is useful to add a penalty term to account for the complexity of the entire weights ($m$) to the Eq. \ref{loss},
\begin{equation}\label{penalty}
    \Delta_\textrm{p} = \frac{\alpha}{2s} \sum_{i=1}^{m} (\boldsymbol{w}^i)^2
\end{equation}
where $\alpha$ is a dimensionless number that controls the degree of penalty. Adding such penalty function in the context of machine learning is called regularization. Then, the optimum solution can be solved by finding the $\boldsymbol{w}$ leading to the zero partial derivative of $\Delta$ with respect to each element in $\boldsymbol{w}$. Accordingly, we use the \textit{numpy.linalg.lstsq}\cite{oliphant2006guide} solver for generalized linear regression problems.

\subsubsection{Neural Network Regression}\label{section:NNP}

\begin{figure}[ht]
\centering
\includegraphics[width=0.45\textwidth]{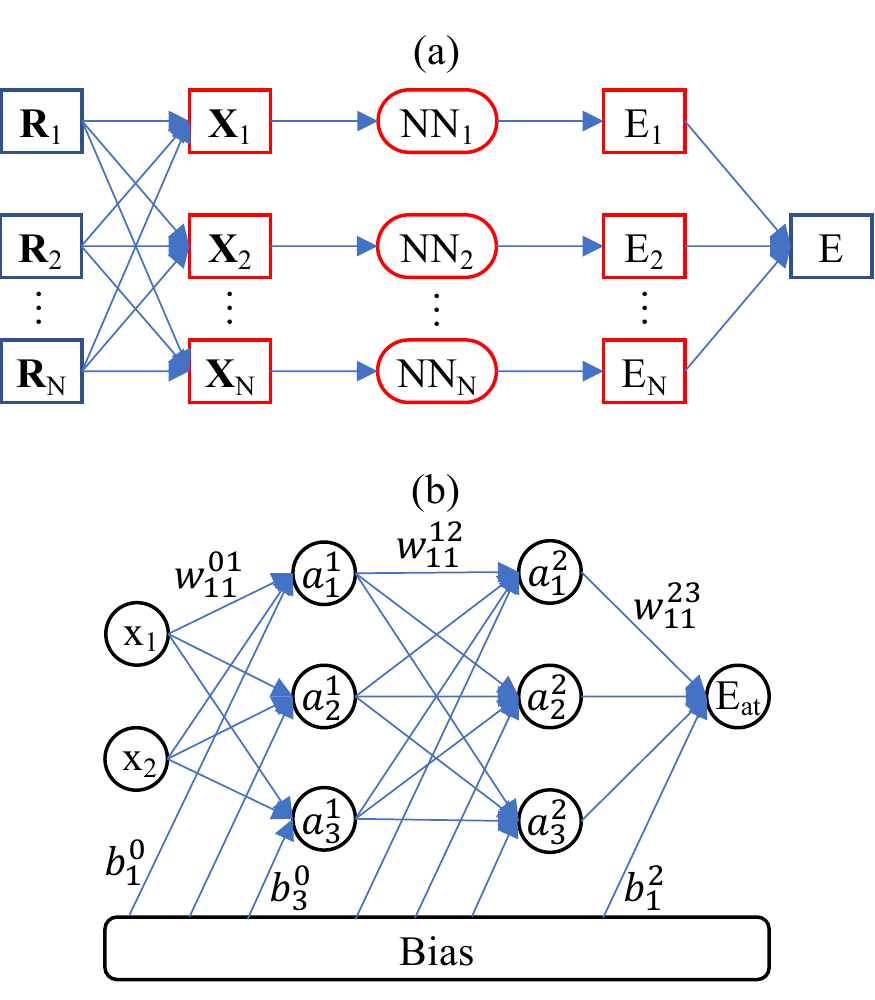}
\caption{(a) A schematic diagram of high-dimensional neural network. The red-colored diagrams are parts of (b) the neural network architecture. Each atom in a structure is firstly mapped into atom-centered descriptors according to the atomic environment of the structure. The atom-centered descriptors serve as the inputs in the neural network architecture that outputs the atomic energy. Finally, the collection of the atomic energies is the total energy of the structure.}
\label{HDNN}
\end{figure}

In this section, the high-dimensional NN (Fig. \ref{HDNN}) is introduced. The regression based on NN can be considered as an extension of linear regression model. For a crystal structure that consists of N atoms, there are N positions ($\boldsymbol{R}_\textrm{N}$) for the atoms to arrange themselves. N atom-centered descriptors ($\boldsymbol{X}_i$) for the structure can be mapped based on this atomic configuration. Each of the atom-centered descriptors is, then, fed into a NN architecture (Fig. \ref{HDNN}b). NN architecture consists of input, hidden, and output neurons. These neurons are organized in layers as shown. The neurons in the first layer (input layer) are occupied by the atom-centered descriptors. The neuron at the output layer defines the atomic energy, $E_i$. Hidden layers lie between the input and output layers. In the case of Fig. \ref{HDNN}b, there are two hidden layers. In particular, we will call this NN architecture, 2-3-3. 2 represents two neurons in the input layers. 3-3 represents two hidden layers with 3 neurons each. It is redundant to repeatedly mention the output layer as the node is always 1. The neurons in hidden layers represent no physical meaning. They act as a functional form to predict the atomic energy. There is no limit to the number of hidden layers. However, the flexibility of NNP will depend on the number of neurons present in the NN architecture. The connectivity in between the neurons are the weight parameters (fitting parameters). Mathematically, one can calculate the value of a neuron in this form:
\begin{equation}\label{neuron}
    X^{l}_{n_i} = a^{l}_{n_i}\bigg( b^{l-1}_{n_i} + \sum^{N}_{n_j=1} W^{l-1, l}_{n_j, n_i} \cdot X^{l-1}_{n_j} \bigg)
\end{equation}
The value of a neuron ($X_{n_i}^l$) at layer $l$ can determined by the relationships between the weights ($W^{l-1, l}_{n_j, n_i}$), the bias ($b^{l-1}_{n_i}$), and all neurons from the previous layer ($X^{l-1}_{n_j}$). $W^{l-1, l}_{n_j, n_i}$ specifies the connectivity of neuron $n_j$ at layer $l-1$ to the neuron $n_i$ at layer $l$. $b^{l-1}_{n_i}$ represents the bias of the previous layer that belongs to the neuron $n_i$. These connectivity are summed based on the total number of neurons ($N$) at layer $l-1$. Finally, an activation function ($a_{n_i}^l$) is applied to the summation to induce non-linearity to the neuron ($X_{n_i}^l$). $X_{n_i}$ at the output layer is equivalent to an atomic energy, and it represents an atom-centered descriptor at the input layer. Since the atomic energy has no reference value to the DFT energy, each atomic energy is collected as in Eq. \ref{total_E} to obtain the total energy of a crystal structure. The accuracy of NNP will rely on the accuracy of the NN architecture to predict the energy.

To train the NNP, we can consistently use the cost function in Eqs. \ref{loss} and \ref{penalty}. The minimization problem is then solved by our in-house stochastic gradient decent and ADAM \cite{Adam} optimizer. Alternatively, we interfaced our in-house code with the Scipy package \cite{scipy}, so it is possible to use the L-BFGS method \cite{L-BFGS} for this study.

\begin{figure*}[ht]
\centering
\includegraphics[width=\textwidth , height=4.5in]{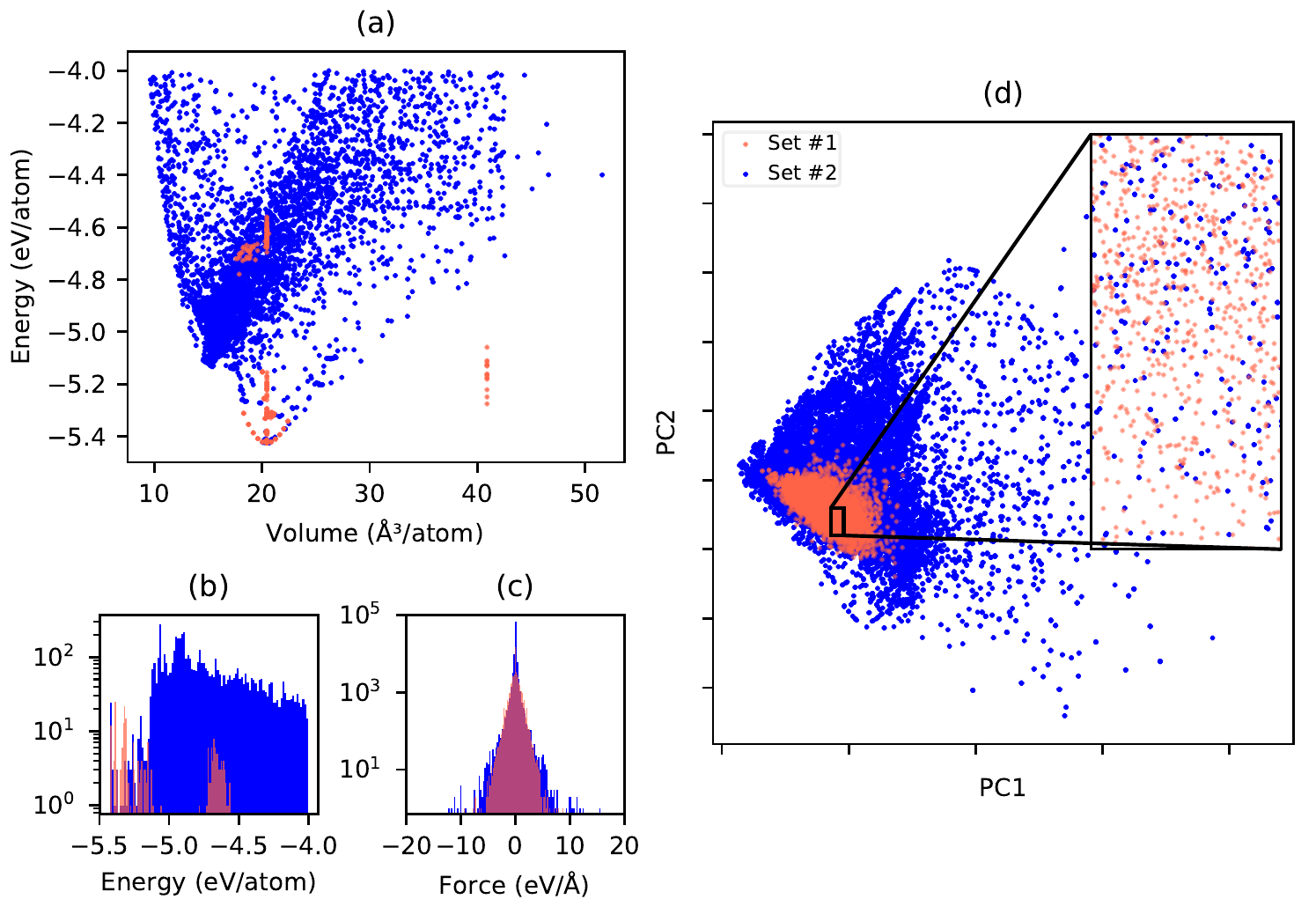}
\caption{(a) The energy versus volume plot for training Set \#1 and Set \#2. The histograms of energy and forces are presented in (b) and (c), respectively. (d) The projection of two most dominating principal components of the atomic bispectrum coefficients. The inset illustrates a zoom-in view of the concentrated area. In the area, Set \#1 is highly concentrated, whereas Set \#2 is more widely spread.}
\label{PES}
\end{figure*}

\section{RESULTS}
\label{sec:results}
In this section, we discuss about the development of accurate and transferable MLFF. First, we introduce two types of data sets---a localized data set and a diverse data set. Second, we will validate our machine learning framework with the localized data set as the baseline. Third, we explore the interplay between bispectrum coefficients and the two machine learning regressions (generalized linear regression and NN) on the localized data set. This subsection is dedicated to further validate the localized data set with a new NNP fitting strategy. Finally, we will develop a transferable silicon MLFF based on the new strategy.

\subsection{Data Sets} \label{pre-training}
Here, we present two silicon data sets. The Set \#1 is the localized data set, obtained from Ref. \cite{zuo2019performance}. Set \#1 contains 244 structures in total (219 for training and 25 for test), which includes the ground state of crystalline structure, strained structures, slabs, and configurations from MD simulations. To generate the diverse data set, we utilized our in-house PyXtal code \cite{pyxtal} to produce thousands of silicon structures with various numbers of atoms in the unit cell from 1, 2, 4, 6, 8 to 16. Random space group (1 to 230) assignment was applied to these silicon structures. For each random structure, we performed four consecutive geometry optimization steps at the level of DFT with steady increase in precision. The maximum numbers for each ionic step were 10, 25, 50 and 50. The relaxed images were then selected to our training pool to represent the shape of PES towards to the energy minima. With this scheme, we ensure that not only the minima, but also the configurations around the minima will be captured during the energy fitting. Afterwards, we performed single-point DFT calculations for all configurations in the training pool using the parameters described in section \ref{setting}. Finally, 5352 silicon structures (Set \#2) were selected by removing structures with energies that are higher than -4.0 eV/atom. In total, Set \#1 has 15078 atoms, and Set \#2 has 31004 atoms. We note that the energy cutoff (600 eV) used in our DFT calculation is slightly higher than the one (520 eV) used in Ref. \cite{zuo2019performance}. However, this resulted in negligible differences according to our test for the same structures. Therefore, we will use these two data sets for direct comparison in the next sections.

As shown in Fig. \ref{PES}, Set \#2 covers more diverse atomic environments in terms of energy, force, and density. Set \#1 includes 244 structures that span from -4.560 to -5.425 eV/atom in energy, and 17.56 to 40.89 $\textrm{\AA}^3$/atom in density. The energy of Set \#2 ranges from -4.0 to -5.425 eV/atom, and the density ranges from 8.295 to 52.81 $\textrm{\AA}^3$/atom. The force distribution in Set \#2 is wider than that in Set \#1. To ensure indirect involvement of Set \#1 to Set \#2, we assessed the data sets by mapping the structures onto the atomic bispectrum coefficients and performed principal component analysis (PCA) on the Set \#2. Then, the atomic bispectrum coefficients of Set \#1 are transformed onto the fitted PCA. The inset shows rare overlapping events between the two data sets in a concentrated area. The data points of Set \#1 cover mostly the empty space in the concentrated area. 

It is important to note that these two sets of data were obtained through entirely different approaches. Clearly, \#2 covers more energy basins in the PES since it was obtained from an unbiased and more uniform sampling. On the other hand, \#1 represents the zoomed region in PES around the equilibrium (i.e., the ground state silicon structure). An MLFF with better coverage of the PES landmarks is useful for an accurate modeling of rare events under various conditions (e.g., phase transitions, pronounced deformations, and chemical reactions). However, many material simulations such as MD are focusing on the region near the equilibrium. As we will discuss in the following sections, fitting \#2 is much more challenging than \#1. While many relatively simple models can yield rather satisfactory errors for \#1, the overall accuracy for \#2 is notably lower regardless whatever methods are applied. Therefore, our goal of this work is to fit a Si-MLFF which can describe \#2 reasonably well while retaining a similar level of accuracy for \#1. 

\begin{table}[b]
  \caption{The setting used to compute the atom-centered descriptors in this study. The Behler-Parrinello descriptors are consistent with Ref. \cite{zuo2019performance}, except that $R_c$ was set to 4.8 {\AA} for the quadratic regression in the previous literature. Moreover, we considered bispectrum coefficients with the band limit $l_\text{max}$ up to 8. The asterisk symbol denotes the reduced parameter set for Behler-Parrinello descriptors.}
\begin{tabular}{ccl}
\hline\hline
Descriptors            & ~~~Parameters~~~     & Values              \\\hline
\multirow{3}{*}{$G^2$} & $R_c$ ({\AA})        & 5.2                     \\
                       & $R_s$ ({\AA})        & 0                        \\
                       & $\eta$ ({\AA}$^{-2}$)& 0.036*, 0.071*, 0.179*, 0.357*, 0.714*,\\
                       &                      & 1.786*, 3.571, 7.142, 17.855  \\\hline
\multirow{4}{*}{$G^4$} & $R_c$ ({\AA})        & 5.2                               \\
                       & $\lambda$ ({\AA})    & -1, 1                             \\
                       & $\zeta$              & 1                                 \\
                       & $\eta$ ({\AA}$^{-2}$)& 0.036*, 0.071*, 0.179*, 0.357*, 0.714,\\
                       &                      & 1.786, 3.571, 7.142, 17.855  \\\hline
\multirow{3}{*}{$B$}   &$R_c$ ({\AA})           & 4.9              \\
                       & $l_\text{max}$       & 2, 3, 4, 5, 6, 7, 8 \\
                       & Normalization        & True, False \\

\hline\hline
\end{tabular}
\label{table:parameters}
\end{table}

To compute the descriptors, we employed the same parameter setting as reported in Ref. \cite{zuo2019performance}, which is summarized in Table \ref{table:parameters}. In the original literature, there were 9 $G^2$ and 18 $G^4$ descriptors. We made a deeper inspection on the histogram of the computed symmetry functions of the entire Set \#1. We identified that descriptors with large $\eta$ values span in a very narrow range. Narrow-range descriptors were less likely to discriminate different local atomic environments, and they could introduce numerical noise. Therefore, we reduced the parameter set, which included only 6 $G^2$ and 8 $G^4$ descriptors for this study. The reduced parameter sets are marked with asterisk symbol. For convenience, we are naming the full Behler-Parrinello descriptors as G27 and the reduced Behler-Parrinello descriptors as G14. For bispectrum coefficient, the expansion is limited to several finite orders, since the higher indices of $l$ can only be beneficial in detecting subtle signals on the neighbor density map. In this study, we only considered the band limit ($l_\text{max}$) up to 8, with focus on 3, 4, and 5 (30, 55, and 91 bispectrum coefficients). They are denoted as B30, B55, and B91. Furthermore, we investigated the case of $B$ with normalization, and they are denoted as $\hat{\text{B}}$30, $\hat{\text{B}}$55, and $\hat{\text{B}}$91.

\subsection{Validation with the Localized Data Set}\label{validation}
\begin{figure}[t]
\centering
\includegraphics[width=0.45\textwidth]{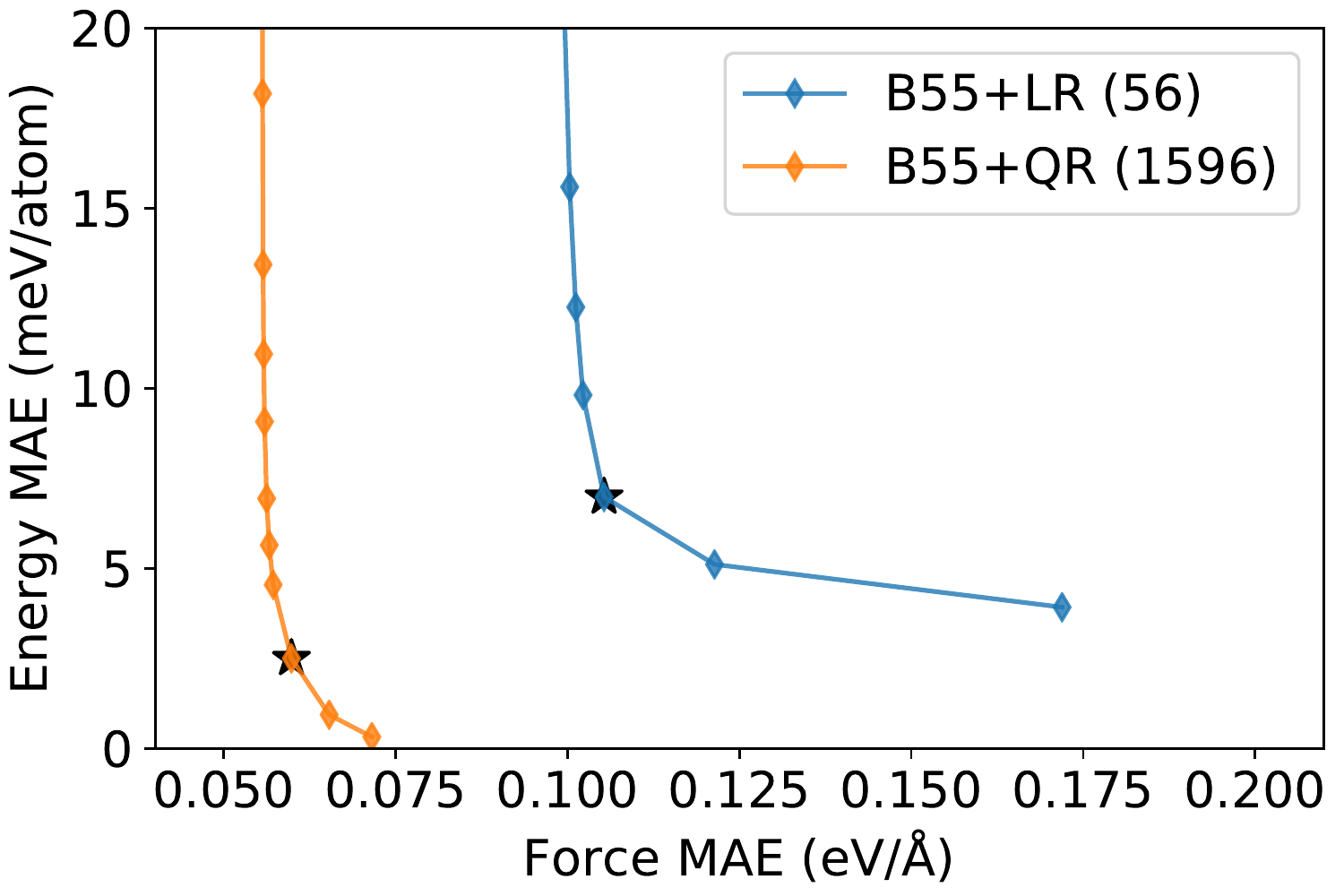}
\caption{The comparison of fitting between linear and quadratic regression based on the B55 descriptors ($l_\text{max}=4$) applied to Set \#1. For each regression, the energy MAE and force MAE values were collected by gradually varying the force coefficients from 1e-6 to 1. The numbers of weight parameters are given in the parentheses. The marked black asterisks correspond the results when the force coefficient is at 1e-4.}
\label{LR-QR}
\end{figure}
In Ref. \cite{zuo2019performance}, the authors presented an extensive benchmark for silicon (as well as several other elemental systems) with different MLFF approaches. This provided us a foundation to validate our MLFF implementations. With Set \#1, we attempted to reproduce the results based on the NNP, SNAP, and quadratic SNAP (qSNAP) methods. They corresponded to the NN regression with G27 descriptors (NNP+G27), linear regression with B55 descriptors (LR+B55), and quadratic regression with B55 descriptors (QR+B55). For the cases of linear and quadratic regressions, the results are deterministic as long as the force coefficient in Eq. \ref{loss} is given. Fig. \ref{LR-QR} displays the gradual changes of mean absolute error (MAE) values for energy and forces by varying the force coefficient ($\beta$) from 1e-6 to 1e+0 for both LR+B55 and QR+B55. For each regression, these points seem to form a Pareto front. Namely, there is no single point which can beat the other points in both energy and force MAE values. Here, we choose a range from the Pareto front which leads to an approximately even change on other sides. This point corresponds to the force coefficient closest to 1e-4. When $\beta$=1e-4, B55+LR yields the MAE values of 6.94 (6.28) meV/atom for energy and 0.11 (0.12) eV/{\AA} for force in training (test) data set. For B55+QR, the results gain significant improvement. The final energy MAE value is 2.50  (2.21) meV/atom, and the force MAE value is 0.06 (0.08) eV/{\AA}. The results are expected since the quadratic form allows the coupling of bispectrum coefficients \cite{wood2018extending}. However, the number of weight parameters also increases notably from 56 to 1596, which increases the computational cost for both FF training and prediction.

\begin{table}[b]
  \caption{The comparison of mean absolute error (MAE) values between this work and Ref. \cite{zuo2019performance} for the same 244 Si data set (Set \#1). The results from Ref. \cite{zuo2019performance} are shown in parentheses. For LR+B55 and QR+B55, the results are shown when force coefficient is at 1e-4. For the NNP fitting, we used NN architectures of 27-24-24 and 14-12-12.}
  \begin{tabular}{ccccc}
    \hline\hline
     Fitting & Train Energy  & Test Energy & Train Force & Test Force \\
    Method        & (meV/atom)    & (meV/atom)  & (eV/\AA) & (eV/\AA) \\
    \hline
    LR+B55   & 6.94 (6.38) & 6.28 (6.89)  &  0.11 (0.21) & 0.12 (0.22) \\
    QR+B55   & 2.50 (3.98) & 2.21 (3.81)  & 0.06 (0.18) & 0.08 (0.17) \\
    NNP+G27  & 5.65 (5.88) & 5.60 (5.60)  & 0.09 (0.12) & 0.11 (0.11) \\
    NNP+G14  & 5.95        & 6.33         & 0.10        & 0.11 \\
    \hline\hline
  \end{tabular}
  \label{ff-results}
\end{table}

\begin{figure}[t]
\centering
\includegraphics[width=0.40\textwidth]{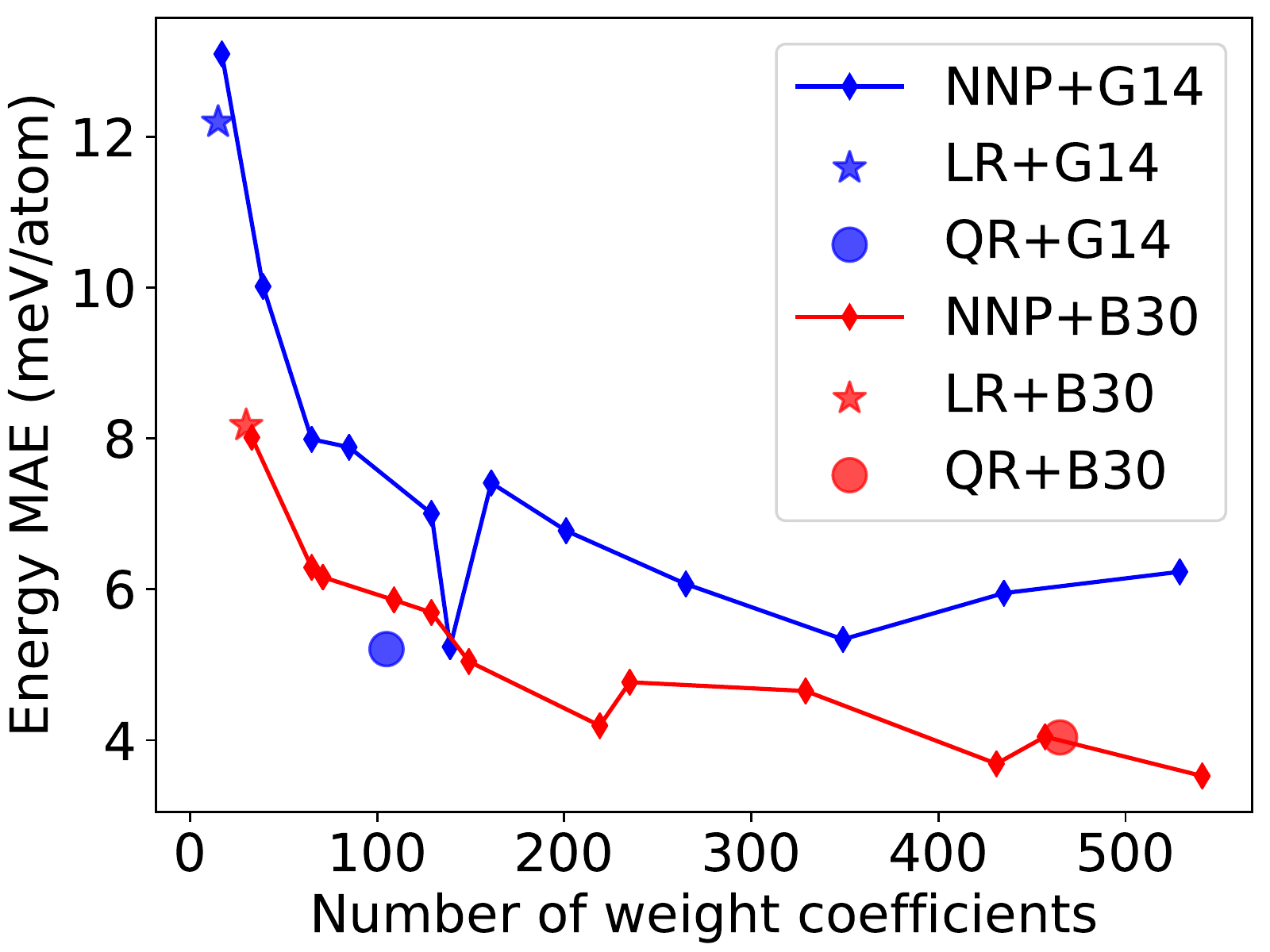}
\caption{The performance of NN regression on G14 and B30 as a function of weight parameters. For comparison, the results from linear and quadratic regressions are also included.}
\label{NNP+G14+B30}
\end{figure}

For NNP+G27, we tested the NNP fitting with NN architecture of 27-24-24. The predicted MAE values are 5.65 meV/atom in the training data set and 5.60 meV/atom in the test data set. The metrics are close to the previously reported values: 5.88 and 5.60 meV/atom in Ref. \cite{zuo2019performance}. Our force MAE values are 0.095 and 0.106 eV/{\AA}, agreeing with the previous report as well. Furthermore, we employed reduced Behler-Parrinello descriptors to the NNP fitting (NNP+G14). We found that the training with NNP+G14 also yielded comparable metrics. This indicated that the removed Behler-Parrinello descriptors descriptors were indeed redundant, and they can cause numerical noise during the NNP training. Correspondingly, we adjusted our NNP training strategy toward G14 to investigate the impacts of hyperparameters on NNP training. In contrast to linear regression, the NNP training is much less vulnerable to the choice of force coefficient since the NNP can compromise for more flexible functional forms. It is rather reliant to the hidden layer size. Fig. \ref{NNP+G14+B30} shows the energy MAE values scanning across the hidden layer sizes for NNP+G14 with $\beta$ fixed at 0.03. Overall picture suggests that NNP performances tend to improve as the NNP model becomes more flexible. However, the NNP accuracy will saturate at some point. Beyond the saturation point, increasing the hidden layer size will only raise the computational cost and lower the chance of finding optimal weight parameters. We also mention that the results from QR+B14 yields better performance than NNP with the same number of parameters. In principle, NNP should be able to self-learn a model similar to QR with the same number of weight parameters. However, different NNP trainings from different initial random guesses may yield somewhat less optimal solutions. This practice suggests that quadratic regression can be an alternative approach when the descriptor size is relatively small.

The results of validation with different training strategies are summarized in Table \ref{ff-results}. Compared to Ref. \cite{zuo2019performance}, our results are close or maybe slightly better, especially in the force performances for generalized linear regression. Therefore, we proceed to make further investigations on Set \#1 by using different strategies.

\subsection{Bispectrum Coefficients/Algorithms Interplay}\label{new_strategy}
\begin{figure}[t]
\centering
\includegraphics[width=0.40\textwidth]{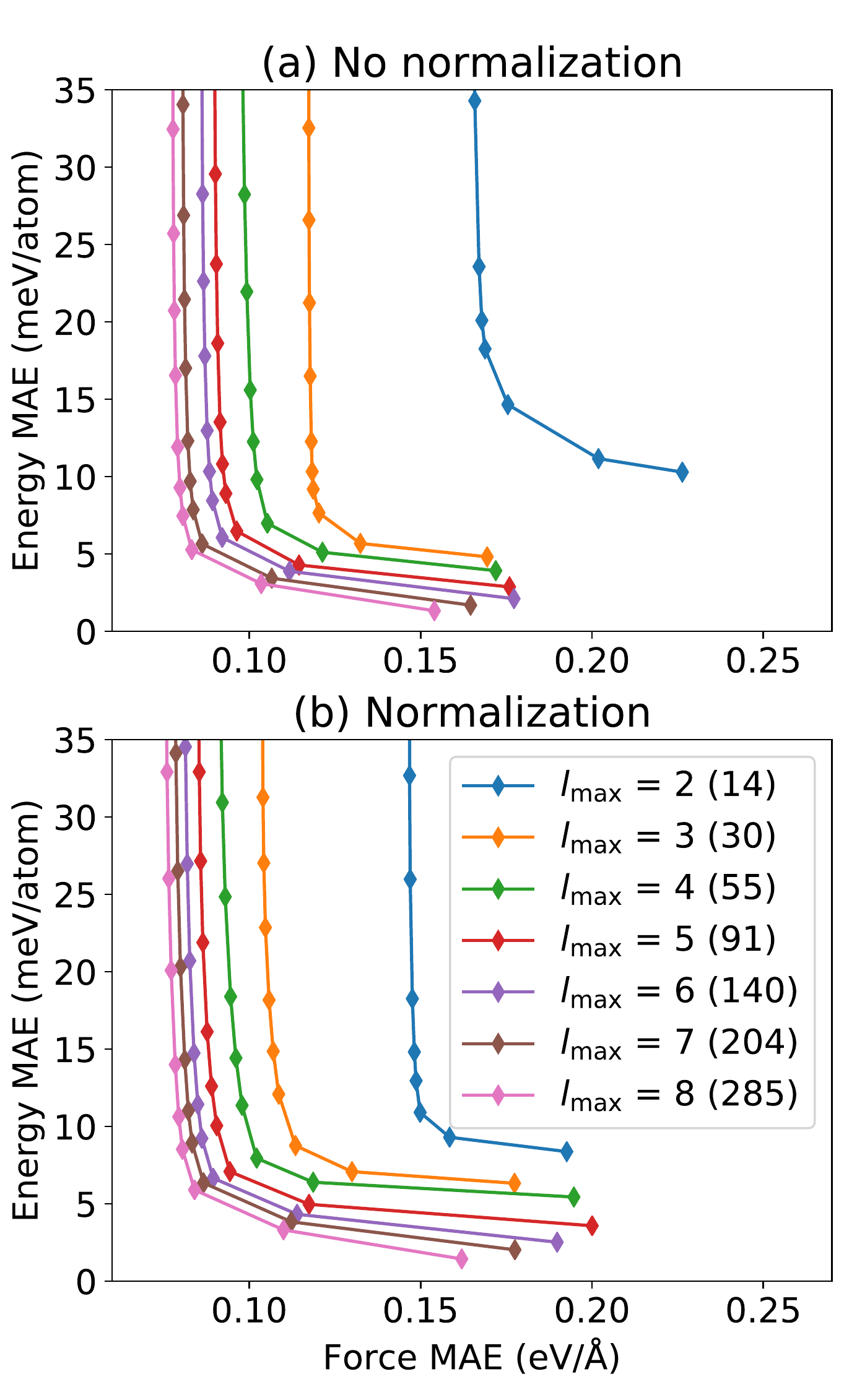}
\caption{The performance of linear regression based on the bispectrum coefficients without (a) and with normalization (b). In each plot, $l_\text{max}$ values from 2 to 8 were considered. The number of descriptors are given in the parenthesis.}
\label{LR-comparison}
\end{figure}
In this section, MLFF fitting with bispectrum coefficients will be discussed in details by using both generalized linear and NN regressions on Set \#1. First, the performances of generalized linear regression can be improved based on the normalization factor of bispectrum coefficients prior to the MLFF fitting. In the original implementation of SNAP \cite{thompson2015spectral}, the bispectrum coefficients are not normalized prior to the MLFF fitting. However, Fig. \ref{LR-comparison} shows the benefits of normalization prior to the MLFF fitting. Linear regression achieves better performances for both energy and forces as $l_\text{max}$ increases. At $l_\text{max} > 5$, there are no significant gains in the MAE values as the computational cost increases. The insignificance of normalization can be due to the limitation of linear regression ability to express the complexity.

Second, Fig. \ref{NNP+G14+B30} shows the overall NNP fitting with bispectrum coefficients as the inputs to the neural network architecture. The results of NNP+B30 are trained with different hidden layer sizes. The best accuracy is achieved with the hidden layer size of [24, 24]. The 30-24-24 architecture consists of 1369 parameters in total. The training MAE values are 3.18 meV/atom and 0.07 eV/\AA, and the test MAE values are 3.54 meV/atom and 0.08 eV/\AA. These metrics reach comparable values to that from QR+B55 (see Table \ref{ff-results}) with less bispectrum coefficients. For reference, linear regression and quadratic regression results with the corresponding number of bispectrum coefficients are also marked in Fig. \ref{NNP+G14+B30}. NNP with bispectrum coefficients can gain notable improvements in comparison to linear regression and quadratic regression. The improvements are expected since NN allows more flexible functional forms to describe the deviation from linearity. Meanwhile, quadratic regression achieves significant improvement in accuracy compared to linear regression due to the extended polynomial forms. However, similar accuracy can be attained with NNP fitting with smaller number of weight parameters.

\subsection{Transferability of the MLFF from a Localized Data Set}\label{inverse}

Our in-house code has the ability to apply various descriptors and regression techniques to train MLFF with satisfactory accuracy ($<$10 meV/atom in energy MAE and $<$0.15 eV/{\AA}~ in force MAE) on Set \#1. From computational perspective, %NNP fitting with Behler-Parrinello descriptors expends less cost than with bispectrum coefficients. However, 
bispectrum coefficients can cover more orthogonal sets and are easier to be expanded. Therefore, we focus on the use of bispectrum coefficients as the main descriptors from now on. Using the MLFF trained on Set \#1, we tried to validate the prediction power on Set \#2 (the more diverse data set). The models include NNP with 30-10-10 architecture (431 parameters, with $\beta$ at 0.03), linear regression (31 parameters), and quadratic regression (528 parameters). The three scenarios use normalized bispectrum coefficients with $l_\text{max}$ of 3, as normalized bispectrum coefficients suggest slight accuracy improvement. Table \ref{nnp-inverse} summarizes the results. In general, the prediction power of the MLFF on Set \#2, especially in energy, is still poor, though the force errors are acceptable. It is not surprising as the machine learning ability in extrapolation is known to be poor. The performance of the MLFF yields great accuracy based on the given training data set. The characteristic of atomic environments of Set \#2 is too broad and most of the data points lay outside of the Set \#1. Therefore, the predicted energy and force are no longer reliable.

Despite the unsatisfactory accuracy, some insights can be gained from this numerical experiment. NN regression can achieve better transferability in comparison to linear and quadratic regression. Although the quadratic regression yields the best accuracy in training (3.99 meV/atom in energy and 0.08 eV/{\AA}) in force, it also produces the largest error on the test set. On the contrary, NN regression achieves a similar level of accuracy on the training (4.70 meV/atom in energy and 0.08 eV/{\AA}). But the errors on the test set (69.8 meV/atom energy MAE and 0.13 eV/{\AA} force MAE) are much smaller. This can be partially explained by the fact that NN adopts more flexible functional forms during fitting.

\begin{table}[ht]
  \caption{The MAE values of the predicted energy and forces of Set \#2 by training on Set \#1. The 30-10-10 is used as the NN architecture for providing comparable weight parameters as the quadratic regression. The numbers inside parentheses are the test MAEs.}
  %\begin{tabular}{lccc}
  \begin{tabular}{l p{1.8cm} p{1.5cm} p{1.5cm}}
    \hline\hline
            & NN & LR  & QR \\
    \hline
    Energy (meV/atom) & 4.7 (70) & 7.5 (110) & 4.0 (265) \\
    Force (eV/\AA) & 0.08 (0.13) & 0.12 (0.15) & 0.08 (0.21) \\
    Number of parameters & 431 & 31 & 496 \\
    \hline\hline
  \end{tabular}
  \label{nnp-inverse}
\end{table}

\subsection{Training with a More Diverse Data Set}\label{SiPyXtal}
%\begin{figure}[t]
%\centering
%\includegraphics[width=0.45\textwidth]{images/NNP+B30.pdf}
%\caption{The performance of NN regression on B30 applied to Set \#1 as a function of %hidden layer size. The following hidden layer sizes were used: [1], [2], [4], [2, 2], %[4, 4], [6, 4], [6, 6], [8, 8], [10, 10], [12, 6], [12, 12], [16, 16], [24, 12], and %[24, 24]. For comparison, the results from linear and quadratic regressions are also %included}
%\label{NNP+B30}
%\end{figure}

\begin{figure}[t]
\centering
\includegraphics[width=0.48\textwidth]{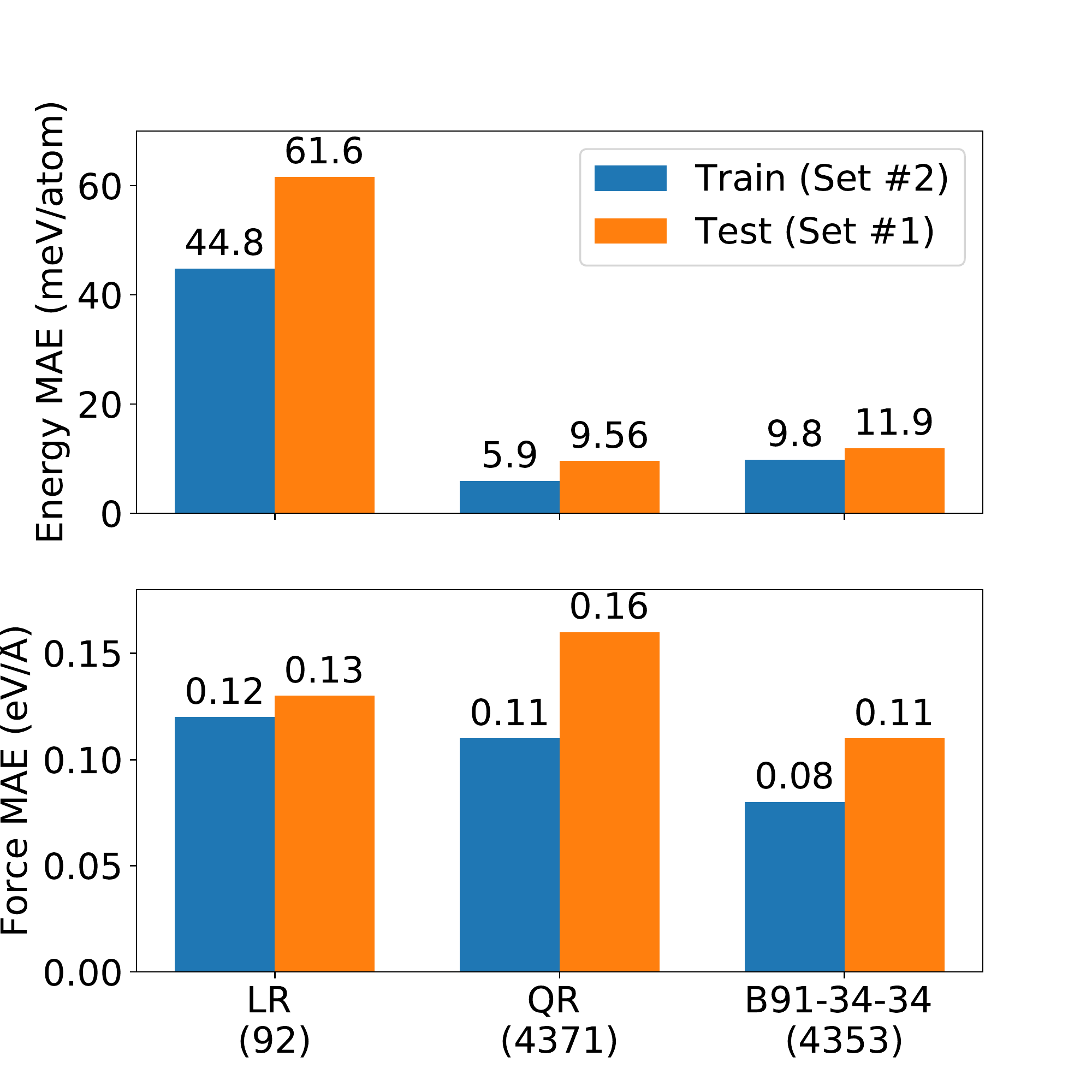}
\caption{The performance of trained MLFF on Set \#2.}
\label{NNP-Set2}
\end{figure}

For the sake of data diversity, it is more natural to train the MLFF based on Set \#2, and test its performance on Set \#1. To train reliable MLFF on Set \#2, we decided to use more bispectrum coefficients and a larger NN architecture and test on Set \#1. In addition, polynomial fittings were included again for the purpose of comparison. For polynomial regression, $l_\text{max}$ at 5 with cutoff radius of 4.9 {\AA} was applied. According to Fig. \ref{LR-comparison}, normalizing the bispectrum coefficients had negligible effect on the results. Hence, normalization was ignored. The $\beta$ value was fixed at 1e-4 for quadratic regression, and 1e-3 for linear regression. The NN architecture of 91-34-34 was used to give a comparable weight parameters as the quadratic regression.

Fig. \ref{NNP-Set2} summarizes the results of Set \#2 training. %The MAE values suggest that the MLFFs are considered transferable. 
In term of energy, quadratic regression performs the best accuracy (5.90 meV/atom), whereas NNP can predict less accurate energy (9.81 meV/atom) but better forces (0.08 eV/{\AA}). It should be emphasized that Set \#2 contains smaller unit cell (1-16 atoms in a unit cell) than Set \#1 (up to 64 atoms in a unit cell). This transition from smaller to larger cells can introduce long-range effects that were not accounted for in the training \cite{kuritz2018size}. Therefore, the MAE values on the test set are consistently larger than the training set. Furthermore, Set \#1 may contain some manually-selected atomic configurations. These configurations may not be fully covered by our random generated structures. While linear regression predicts well on the forces, it guides the energy predictions to unsatisfactory results. This may due to the limitation of the regression technique as the smaller number of parameters fail to describe the true PES. Therefore, our recommendation is to use either quadratic regression (similar to the recently proposed qSNAP method \cite{wood2018extending}) or NN for a better fitting of a diverse dataset. Compared to the quadratic regression, NN is our preferred choice due to its flexibility.  
%However, it should be emphasized that linear regression can provide satisfactory versatility in predicting the physical properties such as elastic constants in the later section.

In comparison to literature, our study yields comparable results as a previous study of diverse silicon cluster \cite{Bartok-PRB-2013}. The authors juxtaposed among several atom-centered descriptors, including bispectrum coefficients, that were coupled with Gaussian process regression. In particular, the root mean square errors (RMSEs) for energy and forces with $l_\text{max}$ at 5 are 20.2 meV/atom and 0.25 eV/{\AA}. Meanwhile, the training RMSEs of our quadratic regression yield 9.7 meV/atom and 0.22 eV/{\AA} and the training RMSEs of 91-34-34 architecture are 14.8 meV/atom and 0.16 eV/{\AA}. In another study, Kuritz \textit{et al.} focuses on training atomic forces using deep learning model with the environmental distances as the descriptors. The force predictions are performed at a scaling from 16 atoms to 128 atoms yields MAE of 0.12 eV/{\AA} \cite{kuritz2018size}, given that the NN nodes are in the order of $10^3$/layer. This phenomenon proves that the choice of descriptor can reduce the complexity of MLFF.

\subsection{Physical Properties}\label{Elastic}

One of the critical requirements for MLFFs is to predict basic material properties, including but not limited to lattice parameter, elastic constants, and bulk moduli of diamond cubic Si. To obtain the elastic constants, we computed the stress-strain relation and fitted the relation to a set of linear equations build from the symmetry. For each applied deformation, the geometry of the structure was optimized to gain net force of zero. The summary of the properties is tabulated in Table \ref{elastic-constant}.

\begin{table*}[ht]
  \caption{The experiment elastic constants \cite{schall2008elastic} of cubic-diamond silicon are shown at zero-Kelvin values, while the DFT data are obtained from Ref. \cite{zuo2019performance}. In comparison to the Gaussian approximation potential (GAP) of Si\cite{Bartok-PRX-2018}, the GAP results is shown below. The numbers of weight parameters are displayed in parentheses. EF and EFS stands for energy-force and energy-force-stress training. LR, QR, and NN are linear, quadratic, and neural network regressions, respectively. The NN architecture for Set \#1 is 30-10-10 and 91-34-34 for Set \#2.}
  
  \begin{tabular}{cccc|cccccc|cccccc}
    \hline\hline
            & &   &   & \multicolumn{6}{c|}{Set \#1} & \multicolumn{6}{c}{Set \#2}\\
    \hline
            & Exp & DFT & GAP & EF-LR & EFS-LR & EF-QR & EFS-QR & EF-NN & EFS-NN & EF-LR & EFS-LR & EF-QR & EFS-QR & EF-NN & EFS-NN \\
            &  &  &  & (56) & (56) & (1596) & (1596) & (431) & (431) & (91) & (91) & (4371) & (4371) & (4353) & (4353) \\
    \hline
    a(\AA) & 5.429 & 5.469 & --- & 5.467 & 5.466 & 5.462 & 5.467 & 5.473 & 5.468 & 5.415 & 5.469 & 5.503 & 5.468 & 5.509 & 5.467\\
    $C_{11}$(GPa) & 167 & 156 & 153 & 153 & 151 & 149 & 152 & 157 & 154 & 137 & 167 & 173 & 158 & 167 & 153\\
    $C_{12}$(GPa) & 65 & 65 & 56 & 100 & 62 & 60 & 57 & 96 & 58 & 76 & 73 & 55 & 55 & 128 & 57\\
    $C_{44}$(GPa) & 81 & 76 & 72 & 69 & 70 & 75 & 75 & 66 & 68 & 73 & 85 & 81 & 71 & 43 & 76\\
    $B_{VRH}$(GPa)& 99 & 95 & 89 & 118 & 92 & 90 & 89 & 117 & 90 & 96 & 104 & 94 & 89 & 141 & 89\\
    \hline\hline
  \end{tabular}
  \label{elastic-constant}
\end{table*}

First, it is crucial to validate our code on Set \#1. On the column of Set \#1 in Table \ref{elastic-constant}, the performances of the MLFFs are presented with different training strategies: energy-force linear regression (EF-LR), energy-force-stress linear regression (EFS-LR), energy-force quadratic regression (EF-QR), energy-force-stress quadratic regression (EFS-QR), energy-force NN (EF-NN), and energy-force-stress NN (EFS-NN). All of the training involved bispectrum coefficients as the descriptor. Linear and quadratic regression used bispectrum coefficients with $l_\text{max}$ of 4, whereas NN used $l_\text{max}$ of 3. Here, we used NN architecture of 30-10-10. Moreover, EF were trained with DFT energy and forces only as the reference values, while EFS included the DFT stress information in the training. Without stress involvement, the quadratic regression performances are the closest to the DFT values. Seemingly, linear and NN regressions fail to extrapolate the $C_{12}$. However, the $C_{12}$ values tend to get closer to the DFT with tiny sacrifice in accuracy of $C_{11}$, when stress is involved. %Moreover, the EFS results insinuate underestimation of the physical properties. It may be due to the nature of the DFT training data.

Second, without stress information, linear and quadratic regression are considered to be more transferable in predicting the physical properties on Set \#2. Evidently, linear regression gains no prominent refinement without trade-off between elastic constants as stress information is added. However, the values are the closest to the experimental values. On the other hand, quadratic regression exhibits accuracy boosts in $C_{11}$ and the lattice constants in comparison to the DFT. %Although NNP fails to accurately reproduce $C_{12}$ and $C_{44}$, the NNP also shows major improvement after including stress information during the training of Set \#2. 
As stress training is employed, NNP seems to benefit the most in term of transferability. Consequently, it is crucial to include stress tensors during the training of NNP.

\section{Discussion}
\textit{Training data set}. In general, MLFF lacks extrapolative ability, unlike the traditional force field method. The training data set plays an extremely important role in MLFF development. A more complete data set can grant the trained MLFF with more powerful predictive ability. The use of randomly pre-symmeterized cyrstal structures is able to produce a data set with highly diverse atomic distribution \cite{pyxtal, Lyakhov-CPC-2013, deringer2018data}. In addition, DFT calculations provide the total energy for each configuration, and the MLFF is trained to describe the total energy of a structure. However, it is possible that the MLFF fails to distinguish the atomic energies for a structures \cite{Huang-PRB-2019, Lee-PRM-2019}. Therefore, Set \#2 includes many structures with smaller unit cells to allow for better descriptions to the PES. Hence, this can help the performance in predicting the total energy. Lastly, Set \#2 can be further extended to consist of more variety in atomic environments to enhance the capability of the current NNP. For example, it was shown above that adding stress tensors can help improving elastic constant predictions.

\textit{Descriptors}. As the complexity of a system's PES increases, different atomic descriptors can yield different accuracy in MLFF development \cite{Bartok-PRB-2013}. For instance, thousands of nodes are needed to achieve similar accuracy in NNP transferability \cite{kuritz2018size}, compared to 34 nodes in this study. The key to extract reliable descriptors is by reconstructing the atomic neighbor density function. The expansion of bispectrum coefficients as the descriptor is more straightforward to be applied than the Behler-Parrinello descriptors. Nevertheless, it is important to take account of the relation between computational cost and accuracy in MLFF training. The current MLFF is developed through the reconstruction of neighbor density function, which is described by the Dirac $\delta$ function. The full description of the true neighbor density can only be partially represented by finite spherical harmonics expansion. In addition, it is numerically unstable to compare the differences between two $\delta$ functions. A better design of descriptor uses smooth Gaussian functions to express the atomic neighbor density, as recently developed in SOAP method \cite{Bartok-PRB-2013}. The comparison between SO(4) bispectrum and SOAP descriptors for NNP development will be conducted in the future code development. Moreover, other similar type of descriptors, such moment tensor potential (MTP) \cite{MTP}, will be investigated in the future.

\textit{Fitting scheme.} Linear regression, as the simplest method in curve fitting, has been used in developing several MLFFs \cite{thompson2015spectral, MTP}. In particular, the MTP approach \cite{zuo2019performance} can predict energy and forces with great accuracy while maintaining acceptable computational cost. The advantage of linear regression method lies in its simple algorithm which provides easy and fast computation. Here, we emphasize that by applying normalization to the atom-centered descriptors can help improving the linear regression training. However, linear/quadratic regression can be sensitive to the noise in the data set. We also applied NN regression in this study. In general NN has more flexibility, which can yield better accuracy, in MLFF fitting. Compared to the linear/quadratic regressions, including stress training in NNP is critical to promote the transferability. %However, overfitting often occurs in NN training. Our code considers regularization term in the loss function to prevent overfitting. 
Beside NN, some non-parametric regression techniques, such as Gaussian Process Regression, have also been proved to be efficient in MLFF development \cite{bartok2015g}. However, this is beyond the scope of the current study.

\textit{Applicability}. For the purpose of MD simulation around the equilibrium state, fitting the MLFF with a localized data set generated from MD simulation is, perhaps, sufficient. However, the primary goal of this work is to generate high quality silicon MLFF for a more general purpose, which requires a complete description of PES for a given chemical system. As discussed above, the MLFF trained with the more diverse data set is generally capable of describing the entire PES better. We expect that the MLFF generated in this work can be used to replace DFT simulation in predicting the structures of crystalline silicon, given that similar works have been done in several elemental systems \cite{podryabinkin2019accelerating, deringer2018data}. Yet, one needs to keep in mind that the quality still depends on the coverage of training data set. For instance, additional data is needed to to enable the prediction for surfaces and clusters \cite{Bartok-PRX-2018}. Moreover, the trained MLFF may not be able to describe the high energy configurations well, since Set \#2 only contains structures with energy less than -4.0 eV/atom. It was found that some nonphysical configurations (e.g., short distances and overly clustered) may be favored under high temperature MD simulations. In this case, it is useful to add a few explicit two-body and three-body terms to prevent the nonphysical configurations \cite{Deringer-carbon-PRB}. We will consider the combination of physical and machine learning terms in the training and investigate the applicability.

\section{Conclusions}

In summary, we present a systematic investigation of MLFFs fitting for elemental silicon using our in-house code. The silicon MLFFs are developed by implementing different regression techniques based on Behler-Parrinello and bispectrum coefficients as the descriptors. The MLFFs trained with Set \#1 (the localized data set) can be described accurately in both energy and forces using generalized linear regression and NN based on both descriptor choices. Among the MLFFs, fitting NNP with the bispectrum coefficients is the most favorable option. This is due to the expansion of bispectrum coefficients is more straightforward than Behler-Parrinello descriptors. In addition, NNP provides more flexible framework in which the functional form can be easily adjusted by adding/reducing the size of weight parameters. %However, Set \#1 has poor transferability in predicting Set \#2---the data set that provides more diversity of silicon.
For Set \#2 generated from random symmetric structures, the NNP fitting with bispectrum coefficients achieves accuracy at 9.8 meV/atom for energy and 0.08 eV/{\AA} for force, which is comparable to the current state of arts based on other approaches. 
%In addition, Set \#2 MLFF can adequately describe Set \#1. 
A thorough study on the applicability of Set \#2 silicon MLFF on more challenging simulations such as crystal structure search will be the subject of our future work.

\section*{Acknowledgments}
We acknowledge the NSF (I-DIRSE-IL: 1940272) and NASA (80NSSC19M0152) for financial support. HY is also supported by the Science Graduate Student Research (SCGSR) program, which is administered by the Oak Ridge Institute for Science and Education (ORISE) for the DOE under contract number DE-SC0014664. The computing resources are provided by XSEDE (TG-DMR180040). A portion of this work was performed under the auspices of the U.S. Department of Energy by Lawrence Livermore National Laboratory under Contract DE-AC52-07NA27344.
The authors thank Dr. A. Thompson at Sandia for insightful discussions in the computation of bispectrum coefficients. We also thank the anonymous referees for excellent suggestions during the revision.
%\end{Acknowledgments}

\section*{Data Availability}
The source source and data that support the findings of this study are available on \url{https://github.com/qzhu2017/PyXtal_FF}.

\section*{References}

\bibliography{reference}

%merlin.mbs aipnum4-1.bst 2010-07-25 4.21a (PWD, AO, DPC) hacked
%Control: key (0)
%Control: author (8) initials jnrlst
%Control: editor formatted (1) identically to author
%Control: production of article title (0) allowed
%Control: page (1) range
%Control: year (1) truncated
%Control: production of eprint (0) enabled
\begin{thebibliography}{53}%
\makeatletter
\providecommand \@ifxundefined [1]{%
 \@ifx{#1\undefined}
}%
\providecommand \@ifnum [1]{%
 \ifnum #1\expandafter \@firstoftwo
 \else \expandafter \@secondoftwo
 \fi
}%
\providecommand \@ifx [1]{%
 \ifx #1\expandafter \@firstoftwo
 \else \expandafter \@secondoftwo
 \fi
}%
\providecommand \natexlab [1]{#1}%
\providecommand \enquote  [1]{``#1''}%
\providecommand \bibnamefont  [1]{#1}%
\providecommand \bibfnamefont [1]{#1}%
\providecommand \citenamefont [1]{#1}%
\providecommand \href@noop [0]{\@secondoftwo}%
\providecommand \href [0]{\begingroup \@sanitize@url \@href}%
\providecommand \@href[1]{\@@startlink{#1}\@@href}%
\providecommand \@@href[1]{\endgroup#1\@@endlink}%
\providecommand \@sanitize@url [0]{\catcode `\\12\catcode `\$12\catcode
  `\&12\catcode `\#12\catcode `\^12\catcode `\_12\catcode `\%12\relax}%
\providecommand \@@startlink[1]{}%
\providecommand \@@endlink[0]{}%
\providecommand \url  [0]{\begingroup\@sanitize@url \@url }%
\providecommand \@url [1]{\endgroup\@href {#1}{\urlprefix }}%
\providecommand \urlprefix  [0]{URL }%
\providecommand \Eprint [0]{\href }%
\providecommand \doibase [0]{http://dx.doi.org/}%
\providecommand \selectlanguage [0]{\@gobble}%
\providecommand \bibinfo  [0]{\@secondoftwo}%
\providecommand \bibfield  [0]{\@secondoftwo}%
\providecommand \translation [1]{[#1]}%
\providecommand \BibitemOpen [0]{}%
\providecommand \bibitemStop [0]{}%
\providecommand \bibitemNoStop [0]{.\EOS\space}%
\providecommand \EOS [0]{\spacefactor3000\relax}%
\providecommand \BibitemShut  [1]{\csname bibitem#1\endcsname}%
\let\auto@bib@innerbib\@empty
%</preamble>
\bibitem [{\citenamefont {Lejaeghere}\ \emph {et~al.}(2016)\citenamefont
  {Lejaeghere}, \citenamefont {Bihlmayer}, \citenamefont {Bj{\"o}rkman},
  \citenamefont {Blaha}, \citenamefont {Bl{\"u}gel}, \citenamefont {Blum},
  \citenamefont {Caliste}, \citenamefont {Castelli}, \citenamefont {Clark},
  \citenamefont {Dal~Corso} \emph {et~al.}}]{lejaeghere2016reproducibility}%
  \BibitemOpen
  \bibfield  {author} {\bibinfo {author} {\bibfnamefont {K.}~\bibnamefont
  {Lejaeghere}}, \bibinfo {author} {\bibfnamefont {G.}~\bibnamefont
  {Bihlmayer}}, \bibinfo {author} {\bibfnamefont {T.}~\bibnamefont
  {Bj{\"o}rkman}}, \bibinfo {author} {\bibfnamefont {P.}~\bibnamefont {Blaha}},
  \bibinfo {author} {\bibfnamefont {S.}~\bibnamefont {Bl{\"u}gel}}, \bibinfo
  {author} {\bibfnamefont {V.}~\bibnamefont {Blum}}, \bibinfo {author}
  {\bibfnamefont {D.}~\bibnamefont {Caliste}}, \bibinfo {author} {\bibfnamefont
  {I.~E.}\ \bibnamefont {Castelli}}, \bibinfo {author} {\bibfnamefont {S.~J.}\
  \bibnamefont {Clark}}, \bibinfo {author} {\bibfnamefont {A.}~\bibnamefont
  {Dal~Corso}},  \emph {et~al.},\ }\bibfield  {title} {\enquote {\bibinfo
  {title} {Reproducibility in density functional theory calculations of
  solids},}\ }\href {\doibase 10.1126/science.aad3000} {\bibfield  {journal}
  {\bibinfo  {journal} {Science}\ }\textbf {\bibinfo {volume} {351}},\ \bibinfo
  {pages} {aad3000} (\bibinfo {year} {2016})}\BibitemShut {NoStop}%
\bibitem [{\citenamefont {Berber}, \citenamefont {Kwon},\ and\ \citenamefont
  {Tom{\'a}nek}(2000)}]{berber2000unusually}%
  \BibitemOpen
  \bibfield  {author} {\bibinfo {author} {\bibfnamefont {S.}~\bibnamefont
  {Berber}}, \bibinfo {author} {\bibfnamefont {Y.-K.}\ \bibnamefont {Kwon}}, \
  and\ \bibinfo {author} {\bibfnamefont {D.}~\bibnamefont {Tom{\'a}nek}},\
  }\bibfield  {title} {\enquote {\bibinfo {title} {Unusually high thermal
  conductivity of carbon nanotubes},}\ }\href {\doibase
  10.1103/PhysRevLett.84.4613} {\bibfield  {journal} {\bibinfo  {journal}
  {Phys. Rev. Lett.}\ }\textbf {\bibinfo {volume} {84}},\ \bibinfo {pages}
  {4613} (\bibinfo {year} {2000})}\BibitemShut {NoStop}%
\bibitem [{\citenamefont {Yamakov}\ \emph {et~al.}(2002)\citenamefont
  {Yamakov}, \citenamefont {Wolf}, \citenamefont {Phillpot}, \citenamefont
  {Mukherjee},\ and\ \citenamefont {Gleiter}}]{yamakov2002dislocation}%
  \BibitemOpen
  \bibfield  {author} {\bibinfo {author} {\bibfnamefont {V.}~\bibnamefont
  {Yamakov}}, \bibinfo {author} {\bibfnamefont {D.}~\bibnamefont {Wolf}},
  \bibinfo {author} {\bibfnamefont {S.~R.}\ \bibnamefont {Phillpot}}, \bibinfo
  {author} {\bibfnamefont {A.~K.}\ \bibnamefont {Mukherjee}}, \ and\ \bibinfo
  {author} {\bibfnamefont {H.}~\bibnamefont {Gleiter}},\ }\bibfield  {title}
  {\enquote {\bibinfo {title} {Dislocation processes in the deformation of
  nanocrystalline aluminium by molecular-dynamics simulation},}\ }\href
  {\doibase 10.1038/nmat700} {\bibfield  {journal} {\bibinfo  {journal} {Nat.
  Mater.}\ }\textbf {\bibinfo {volume} {1}},\ \bibinfo {pages} {45} (\bibinfo
  {year} {2002})}\BibitemShut {NoStop}%
\bibitem [{\citenamefont {Yamakov}\ \emph {et~al.}(2004)\citenamefont
  {Yamakov}, \citenamefont {Wolf}, \citenamefont {Phillpot}, \citenamefont
  {Mukherjee},\ and\ \citenamefont {Gleiter}}]{yamakov2004deformation}%
  \BibitemOpen
  \bibfield  {author} {\bibinfo {author} {\bibfnamefont {V.}~\bibnamefont
  {Yamakov}}, \bibinfo {author} {\bibfnamefont {D.}~\bibnamefont {Wolf}},
  \bibinfo {author} {\bibfnamefont {S.}~\bibnamefont {Phillpot}}, \bibinfo
  {author} {\bibfnamefont {A.}~\bibnamefont {Mukherjee}}, \ and\ \bibinfo
  {author} {\bibfnamefont {H.}~\bibnamefont {Gleiter}},\ }\bibfield  {title}
  {\enquote {\bibinfo {title} {Deformation-mechanism map for nanocrystalline
  metals by molecular-dynamics simulation},}\ }\href {\doibase
  10.1038/nmat1035} {\bibfield  {journal} {\bibinfo  {journal} {Nat. Mater.}\
  }\textbf {\bibinfo {volume} {3}},\ \bibinfo {pages} {43} (\bibinfo {year}
  {2004})}\BibitemShut {NoStop}%
\bibitem [{\citenamefont {Oganov}\ \emph {et~al.}(2019)\citenamefont {Oganov},
  \citenamefont {Pickard}, \citenamefont {Zhu},\ and\ \citenamefont
  {Needs}}]{oganov2019structure}%
  \BibitemOpen
  \bibfield  {author} {\bibinfo {author} {\bibfnamefont {A.~R.}\ \bibnamefont
  {Oganov}}, \bibinfo {author} {\bibfnamefont {C.~J.}\ \bibnamefont {Pickard}},
  \bibinfo {author} {\bibfnamefont {Q.}~\bibnamefont {Zhu}}, \ and\ \bibinfo
  {author} {\bibfnamefont {R.~J.}\ \bibnamefont {Needs}},\ }\bibfield  {title}
  {\enquote {\bibinfo {title} {{Structure prediction drives materials
  discovery}},}\ }\href {\doibase 10.1038/s41578-019-0101-8} {\bibfield
  {journal} {\bibinfo  {journal} {Nat. Rev. Mater.}\ }\textbf {\bibinfo
  {volume} {4}},\ \bibinfo {pages} {331–348} (\bibinfo {year}
  {2019})}\BibitemShut {NoStop}%
\bibitem [{\citenamefont {Curtarolo}\ \emph {et~al.}(2013)\citenamefont
  {Curtarolo}, \citenamefont {Hart}, \citenamefont {Nardelli}, \citenamefont
  {Mingo}, \citenamefont {Sanvito},\ and\ \citenamefont
  {Levy}}]{Curtarolo-2013}%
  \BibitemOpen
  \bibfield  {author} {\bibinfo {author} {\bibfnamefont {S.}~\bibnamefont
  {Curtarolo}}, \bibinfo {author} {\bibfnamefont {G.~L.}\ \bibnamefont {Hart}},
  \bibinfo {author} {\bibfnamefont {M.~B.}\ \bibnamefont {Nardelli}}, \bibinfo
  {author} {\bibfnamefont {N.}~\bibnamefont {Mingo}}, \bibinfo {author}
  {\bibfnamefont {S.}~\bibnamefont {Sanvito}}, \ and\ \bibinfo {author}
  {\bibfnamefont {O.}~\bibnamefont {Levy}},\ }\bibfield  {title} {\enquote
  {\bibinfo {title} {The high-throughput highway to computational materials
  design},}\ }\href {\doibase 10.1038/nmat3568} {\bibfield  {journal} {\bibinfo
   {journal} {Nat. Mater.}\ }\textbf {\bibinfo {volume} {12}},\ \bibinfo
  {pages} {191} (\bibinfo {year} {2013})}\BibitemShut {NoStop}%
\bibitem [{\citenamefont {Bart{\'o}k}\ \emph {et~al.}(2013)\citenamefont
  {Bart{\'o}k}, \citenamefont {Gillan}, \citenamefont {Manby},\ and\
  \citenamefont {Cs{\'a}nyi}}]{bartok2013machine}%
  \BibitemOpen
  \bibfield  {author} {\bibinfo {author} {\bibfnamefont {A.~P.}\ \bibnamefont
  {Bart{\'o}k}}, \bibinfo {author} {\bibfnamefont {M.~J.}\ \bibnamefont
  {Gillan}}, \bibinfo {author} {\bibfnamefont {F.~R.}\ \bibnamefont {Manby}}, \
  and\ \bibinfo {author} {\bibfnamefont {G.}~\bibnamefont {Cs{\'a}nyi}},\
  }\bibfield  {title} {\enquote {\bibinfo {title} {Machine-learning approach
  for one-and two-body corrections to density functional theory: Applications
  to molecular and condensed water},}\ }\href {\doibase
  10.1103/PhysRevB.88.054104} {\bibfield  {journal} {\bibinfo  {journal} {Phys.
  Rev. B}\ }\textbf {\bibinfo {volume} {88}},\ \bibinfo {pages} {054104}
  (\bibinfo {year} {2013})}\BibitemShut {NoStop}%
\bibitem [{\citenamefont {Artrith}, \citenamefont {Morawietz},\ and\
  \citenamefont {Behler}(2011)}]{artrith2011high}%
  \BibitemOpen
  \bibfield  {author} {\bibinfo {author} {\bibfnamefont {N.}~\bibnamefont
  {Artrith}}, \bibinfo {author} {\bibfnamefont {T.}~\bibnamefont {Morawietz}},
  \ and\ \bibinfo {author} {\bibfnamefont {J.}~\bibnamefont {Behler}},\
  }\bibfield  {title} {\enquote {\bibinfo {title} {High-dimensional
  neural-network potentials for multicomponent systems: Applications to zinc
  oxide},}\ }\href {\doibase 10.1103/PhysRevB.83.153101} {\bibfield  {journal}
  {\bibinfo  {journal} {Phys. Rev. B}\ }\textbf {\bibinfo {volume} {83}},\
  \bibinfo {pages} {153101} (\bibinfo {year} {2011})}\BibitemShut {NoStop}%
\bibitem [{\citenamefont {Khaliullin}\ \emph {et~al.}(2011)\citenamefont
  {Khaliullin}, \citenamefont {Eshet}, \citenamefont {K{\"u}hne}, \citenamefont
  {Behler},\ and\ \citenamefont {Parrinello}}]{khaliullin2011nucleation}%
  \BibitemOpen
  \bibfield  {author} {\bibinfo {author} {\bibfnamefont {R.~Z.}\ \bibnamefont
  {Khaliullin}}, \bibinfo {author} {\bibfnamefont {H.}~\bibnamefont {Eshet}},
  \bibinfo {author} {\bibfnamefont {T.~D.}\ \bibnamefont {K{\"u}hne}}, \bibinfo
  {author} {\bibfnamefont {J.}~\bibnamefont {Behler}}, \ and\ \bibinfo {author}
  {\bibfnamefont {M.}~\bibnamefont {Parrinello}},\ }\bibfield  {title}
  {\enquote {\bibinfo {title} {Nucleation mechanism for the direct
  graphite-to-diamond phase transition},}\ }\href {\doibase
  https://doi.org/10.1038/nmat3078} {\bibfield  {journal} {\bibinfo  {journal}
  {Nat. Mater.}\ }\textbf {\bibinfo {volume} {10}},\ \bibinfo {pages} {693}
  (\bibinfo {year} {2011})}\BibitemShut {NoStop}%
\bibitem [{\citenamefont {Behler}\ \emph {et~al.}(2008)\citenamefont {Behler},
  \citenamefont {Marto{\v{n}}{\'a}k}, \citenamefont {Donadio},\ and\
  \citenamefont {Parrinello}}]{behler2008metadynamics}%
  \BibitemOpen
  \bibfield  {author} {\bibinfo {author} {\bibfnamefont {J.}~\bibnamefont
  {Behler}}, \bibinfo {author} {\bibfnamefont {R.}~\bibnamefont
  {Marto{\v{n}}{\'a}k}}, \bibinfo {author} {\bibfnamefont {D.}~\bibnamefont
  {Donadio}}, \ and\ \bibinfo {author} {\bibfnamefont {M.}~\bibnamefont
  {Parrinello}},\ }\bibfield  {title} {\enquote {\bibinfo {title} {Metadynamics
  simulations of the high-pressure phases of silicon employing a
  high-dimensional neural network potential},}\ }\href {\doibase
  10.1103/PhysRevLett.100.185501} {\bibfield  {journal} {\bibinfo  {journal}
  {Phys. Rev. Lett.}\ }\textbf {\bibinfo {volume} {100}},\ \bibinfo {pages}
  {185501} (\bibinfo {year} {2008})}\BibitemShut {NoStop}%
\bibitem [{\citenamefont {Thompson}\ \emph {et~al.}(2015)\citenamefont
  {Thompson}, \citenamefont {Swiler}, \citenamefont {Trott}, \citenamefont
  {Foiles},\ and\ \citenamefont {Tucker}}]{thompson2015spectral}%
  \BibitemOpen
  \bibfield  {author} {\bibinfo {author} {\bibfnamefont {A.~P.}\ \bibnamefont
  {Thompson}}, \bibinfo {author} {\bibfnamefont {L.~P.}\ \bibnamefont
  {Swiler}}, \bibinfo {author} {\bibfnamefont {C.~R.}\ \bibnamefont {Trott}},
  \bibinfo {author} {\bibfnamefont {S.~M.}\ \bibnamefont {Foiles}}, \ and\
  \bibinfo {author} {\bibfnamefont {G.~J.}\ \bibnamefont {Tucker}},\ }\bibfield
   {title} {\enquote {\bibinfo {title} {Spectral neighbor analysis method for
  automated generation of quantum-accurate interatomic potentials},}\ }\href
  {\doibase 10.1016/j.jcp.2014.12.018} {\bibfield  {journal} {\bibinfo
  {journal} {J. Comput. Phys.}\ }\textbf {\bibinfo {volume} {285}},\ \bibinfo
  {pages} {316--330} (\bibinfo {year} {2015})}\BibitemShut {NoStop}%
\bibitem [{\citenamefont {Wood}\ and\ \citenamefont
  {Thompson}(2018)}]{wood2018extending}%
  \BibitemOpen
  \bibfield  {author} {\bibinfo {author} {\bibfnamefont {M.~A.}\ \bibnamefont
  {Wood}}\ and\ \bibinfo {author} {\bibfnamefont {A.~P.}\ \bibnamefont
  {Thompson}},\ }\bibfield  {title} {\enquote {\bibinfo {title} {Extending the
  accuracy of the snap interatomic potential form},}\ }\href {\doibase
  10.1063/1.5017641} {\bibfield  {journal} {\bibinfo  {journal} {J. Chem.
  Phys.}\ }\textbf {\bibinfo {volume} {148}},\ \bibinfo {pages} {241721}
  (\bibinfo {year} {2018})}\BibitemShut {NoStop}%
\bibitem [{\citenamefont {Pozdnyakov}\ \emph {et~al.}(2019)\citenamefont
  {Pozdnyakov}, \citenamefont {Oganov}, \citenamefont {Mazitov}, \citenamefont
  {Frolov}, \citenamefont {Kruglov},\ and\ \citenamefont
  {Mazhnik}}]{pozdnyakov2019fast}%
  \BibitemOpen
  \bibfield  {author} {\bibinfo {author} {\bibfnamefont {S.}~\bibnamefont
  {Pozdnyakov}}, \bibinfo {author} {\bibfnamefont {A.~R.}\ \bibnamefont
  {Oganov}}, \bibinfo {author} {\bibfnamefont {A.}~\bibnamefont {Mazitov}},
  \bibinfo {author} {\bibfnamefont {T.}~\bibnamefont {Frolov}}, \bibinfo
  {author} {\bibfnamefont {I.}~\bibnamefont {Kruglov}}, \ and\ \bibinfo
  {author} {\bibfnamefont {E.}~\bibnamefont {Mazhnik}},\ }\bibfield  {title}
  {\enquote {\bibinfo {title} {Fast general two-and three-body interatomic
  potential},}\ }\href@noop {} {\bibfield  {journal} {\bibinfo  {journal}
  {arXiv preprint arXiv:1910.07513}\ } (\bibinfo {year} {2019})},\ \Eprint
  {http://arxiv.org/abs/1910.07513} {arXiv:1910.07513 [physics.comp-ph]}
  \BibitemShut {NoStop}%
\bibitem [{\citenamefont {Shapeev}(2016)}]{MTP}%
  \BibitemOpen
  \bibfield  {author} {\bibinfo {author} {\bibfnamefont {A.~V.}\ \bibnamefont
  {Shapeev}},\ }\bibfield  {title} {\enquote {\bibinfo {title} {Moment tensor
  potentials: A class of systematically improvable interatomic potentials},}\
  }\href {\doibase 10.1137/15M1054183} {\bibfield  {journal} {\bibinfo
  {journal} {Multiscale Model. Simul.}\ }\textbf {\bibinfo {volume} {14}},\
  \bibinfo {pages} {1153--1173} (\bibinfo {year} {2016})}\BibitemShut {NoStop}%
\bibitem [{\citenamefont {Bart{\'o}k}\ \emph {et~al.}(2010)\citenamefont
  {Bart{\'o}k}, \citenamefont {Payne}, \citenamefont {Kondor},\ and\
  \citenamefont {Cs{\'a}nyi}}]{bartok2010gaussian}%
  \BibitemOpen
  \bibfield  {author} {\bibinfo {author} {\bibfnamefont {A.~P.}\ \bibnamefont
  {Bart{\'o}k}}, \bibinfo {author} {\bibfnamefont {M.~C.}\ \bibnamefont
  {Payne}}, \bibinfo {author} {\bibfnamefont {R.}~\bibnamefont {Kondor}}, \
  and\ \bibinfo {author} {\bibfnamefont {G.}~\bibnamefont {Cs{\'a}nyi}},\
  }\bibfield  {title} {\enquote {\bibinfo {title} {Gaussian approximation
  potentials: The accuracy of quantum mechanics, without the electrons},}\
  }\href {\doibase 10.1103/PhysRevLett.104.136403} {\bibfield  {journal}
  {\bibinfo  {journal} {Phys. Rev. Lett.}\ }\textbf {\bibinfo {volume} {104}},\
  \bibinfo {pages} {136403} (\bibinfo {year} {2010})}\BibitemShut {NoStop}%
\bibitem [{\citenamefont {Bart{\'o}k}\ and\ \citenamefont
  {Cs{\'a}nyi}(2015)}]{bartok2015g}%
  \BibitemOpen
  \bibfield  {author} {\bibinfo {author} {\bibfnamefont {A.~P.}\ \bibnamefont
  {Bart{\'o}k}}\ and\ \bibinfo {author} {\bibfnamefont {G.}~\bibnamefont
  {Cs{\'a}nyi}},\ }\bibfield  {title} {\enquote {\bibinfo {title} {Gaussian
  approximation potentials: A brief tutorial introduction},}\ }\href {\doibase
  10.1002/qua.24927} {\bibfield  {journal} {\bibinfo  {journal} {Int. J.
  Quantum Chem.}\ }\textbf {\bibinfo {volume} {115}},\ \bibinfo {pages}
  {1051--1057} (\bibinfo {year} {2015})}\BibitemShut {NoStop}%
\bibitem [{\citenamefont {Behler}\ and\ \citenamefont
  {Parrinello}(2007)}]{behler2007generalized}%
  \BibitemOpen
  \bibfield  {author} {\bibinfo {author} {\bibfnamefont {J.}~\bibnamefont
  {Behler}}\ and\ \bibinfo {author} {\bibfnamefont {M.}~\bibnamefont
  {Parrinello}},\ }\bibfield  {title} {\enquote {\bibinfo {title} {Generalized
  neural-network representation of high-dimensional potential-energy
  surfaces},}\ }\href {\doibase 10.1103/PhysRevLett.98.146401} {\bibfield
  {journal} {\bibinfo  {journal} {Phys. Rev. Lett.}\ }\textbf {\bibinfo
  {volume} {98}},\ \bibinfo {pages} {146401} (\bibinfo {year}
  {2007})}\BibitemShut {NoStop}%
\bibitem [{\citenamefont {Behler}(2015)}]{behler2015constructing}%
  \BibitemOpen
  \bibfield  {author} {\bibinfo {author} {\bibfnamefont {J.}~\bibnamefont
  {Behler}},\ }\bibfield  {title} {\enquote {\bibinfo {title} {Constructing
  high-dimensional neural network potentials: A tutorial review},}\ }\href
  {\doibase 10.1002/qua.24890} {\bibfield  {journal} {\bibinfo  {journal} {Int.
  J. Quantum Chem.}\ }\textbf {\bibinfo {volume} {115}},\ \bibinfo {pages}
  {1032--1050} (\bibinfo {year} {2015})}\BibitemShut {NoStop}%
\bibitem [{\citenamefont {Zuo}\ \emph {et~al.}(2020)\citenamefont {Zuo},
  \citenamefont {Chen}, \citenamefont {Li}, \citenamefont {Deng}, \citenamefont
  {Chen}, \citenamefont {Behler}, \citenamefont {Cs{\'a}nyi}, \citenamefont
  {Shapeev}, \citenamefont {Thompson}, \citenamefont {Wood} \emph
  {et~al.}}]{zuo2019performance}%
  \BibitemOpen
  \bibfield  {author} {\bibinfo {author} {\bibfnamefont {Y.}~\bibnamefont
  {Zuo}}, \bibinfo {author} {\bibfnamefont {C.}~\bibnamefont {Chen}}, \bibinfo
  {author} {\bibfnamefont {X.}~\bibnamefont {Li}}, \bibinfo {author}
  {\bibfnamefont {Z.}~\bibnamefont {Deng}}, \bibinfo {author} {\bibfnamefont
  {Y.}~\bibnamefont {Chen}}, \bibinfo {author} {\bibfnamefont {J.}~\bibnamefont
  {Behler}}, \bibinfo {author} {\bibfnamefont {G.}~\bibnamefont {Cs{\'a}nyi}},
  \bibinfo {author} {\bibfnamefont {A.~V.}\ \bibnamefont {Shapeev}}, \bibinfo
  {author} {\bibfnamefont {A.~P.}\ \bibnamefont {Thompson}}, \bibinfo {author}
  {\bibfnamefont {M.~A.}\ \bibnamefont {Wood}},  \emph {et~al.},\ }\bibfield
  {title} {\enquote {\bibinfo {title} {Performance and cost assessment of
  machine learning interatomic potentials},}\ }\href {\doibase
  https://doi.org/10.1021/acs.jpca.9b08723} {\bibfield  {journal} {\bibinfo
  {journal} {J. Phys. Chem. A}\ }\textbf {\bibinfo {volume} {124}},\ \bibinfo
  {pages} {731--745} (\bibinfo {year} {2020})}\BibitemShut {NoStop}%
\bibitem [{\citenamefont {Hajinazar}, \citenamefont {Shao},\ and\ \citenamefont
  {Kolmogorov}(2017)}]{Hajinazar-PRB-2017}%
  \BibitemOpen
  \bibfield  {author} {\bibinfo {author} {\bibfnamefont {S.}~\bibnamefont
  {Hajinazar}}, \bibinfo {author} {\bibfnamefont {J.}~\bibnamefont {Shao}}, \
  and\ \bibinfo {author} {\bibfnamefont {A.~N.}\ \bibnamefont {Kolmogorov}},\
  }\bibfield  {title} {\enquote {\bibinfo {title} {Stratified construction of
  neural network based interatomic models for multicomponent materials},}\
  }\href {\doibase 10.1103/PhysRevB.95.014114} {\bibfield  {journal} {\bibinfo
  {journal} {Phys. Rev. B}\ }\textbf {\bibinfo {volume} {95}},\ \bibinfo
  {pages} {014114} (\bibinfo {year} {2017})}\BibitemShut {NoStop}%
\bibitem [{\citenamefont {Deringer}, \citenamefont {Pickard},\ and\
  \citenamefont {Cs{\'a}nyi}(2018)}]{deringer2018data}%
  \BibitemOpen
  \bibfield  {author} {\bibinfo {author} {\bibfnamefont {V.~L.}\ \bibnamefont
  {Deringer}}, \bibinfo {author} {\bibfnamefont {C.~J.}\ \bibnamefont
  {Pickard}}, \ and\ \bibinfo {author} {\bibfnamefont {G.}~\bibnamefont
  {Cs{\'a}nyi}},\ }\bibfield  {title} {\enquote {\bibinfo {title} {Data-driven
  learning of total and local energies in elemental boron},}\ }\href {\doibase
  10.1103/PhysRevLett.120.156001} {\bibfield  {journal} {\bibinfo  {journal}
  {Phys. Rev. Lett.}\ }\textbf {\bibinfo {volume} {120}},\ \bibinfo {pages}
  {156001} (\bibinfo {year} {2018})}\BibitemShut {NoStop}%
\bibitem [{\citenamefont {Podryabinkin}\ \emph {et~al.}(2019)\citenamefont
  {Podryabinkin}, \citenamefont {Tikhonov}, \citenamefont {Shapeev},\ and\
  \citenamefont {Oganov}}]{podryabinkin2019accelerating}%
  \BibitemOpen
  \bibfield  {author} {\bibinfo {author} {\bibfnamefont {E.~V.}\ \bibnamefont
  {Podryabinkin}}, \bibinfo {author} {\bibfnamefont {E.~V.}\ \bibnamefont
  {Tikhonov}}, \bibinfo {author} {\bibfnamefont {A.~V.}\ \bibnamefont
  {Shapeev}}, \ and\ \bibinfo {author} {\bibfnamefont {A.~R.}\ \bibnamefont
  {Oganov}},\ }\bibfield  {title} {\enquote {\bibinfo {title} {Accelerating
  crystal structure prediction by machine-learning interatomic potentials with
  active learning},}\ }\href {\doibase 10.1103/PhysRevB.99.064114} {\bibfield
  {journal} {\bibinfo  {journal} {Phys. Rev. B}\ }\textbf {\bibinfo {volume}
  {99}},\ \bibinfo {pages} {064114} (\bibinfo {year} {2019})}\BibitemShut
  {NoStop}%
\bibitem [{\citenamefont {Jacobsen}, \citenamefont {J\o{}rgensen},\ and\
  \citenamefont {Hammer}(2018)}]{Jacobsen-PRL-2018}%
  \BibitemOpen
  \bibfield  {author} {\bibinfo {author} {\bibfnamefont {T.~L.}\ \bibnamefont
  {Jacobsen}}, \bibinfo {author} {\bibfnamefont {M.~S.}\ \bibnamefont
  {J\o{}rgensen}}, \ and\ \bibinfo {author} {\bibfnamefont {B.}~\bibnamefont
  {Hammer}},\ }\bibfield  {title} {\enquote {\bibinfo {title} {On-the-fly
  machine learning of atomic potential in density functional theory structure
  optimization},}\ }\href {\doibase 10.1103/PhysRevLett.120.026102} {\bibfield
  {journal} {\bibinfo  {journal} {Phys. Rev. Lett.}\ }\textbf {\bibinfo
  {volume} {120}},\ \bibinfo {pages} {026102} (\bibinfo {year}
  {2018})}\BibitemShut {NoStop}%
\bibitem [{\citenamefont {Zeni}\ \emph {et~al.}(2018)\citenamefont {Zeni},
  \citenamefont {Rossi}, \citenamefont {Glielmo}, \citenamefont {Fekete},
  \citenamefont {Gaston}, \citenamefont {Baletto},\ and\ \citenamefont
  {De~Vita}}]{zeni2018building}%
  \BibitemOpen
  \bibfield  {author} {\bibinfo {author} {\bibfnamefont {C.}~\bibnamefont
  {Zeni}}, \bibinfo {author} {\bibfnamefont {K.}~\bibnamefont {Rossi}},
  \bibinfo {author} {\bibfnamefont {A.}~\bibnamefont {Glielmo}}, \bibinfo
  {author} {\bibfnamefont {{\'A}.}~\bibnamefont {Fekete}}, \bibinfo {author}
  {\bibfnamefont {N.}~\bibnamefont {Gaston}}, \bibinfo {author} {\bibfnamefont
  {F.}~\bibnamefont {Baletto}}, \ and\ \bibinfo {author} {\bibfnamefont
  {A.}~\bibnamefont {De~Vita}},\ }\bibfield  {title} {\enquote {\bibinfo
  {title} {Building machine learning force fields for nanoclusters},}\ }\href
  {\doibase 10.1063/1.5024558} {\bibfield  {journal} {\bibinfo  {journal} {J.
  Chem. Phys.}\ }\textbf {\bibinfo {volume} {148}},\ \bibinfo {pages} {241739}
  (\bibinfo {year} {2018})}\BibitemShut {NoStop}%
\bibitem [{\citenamefont {Bart\'ok}\ \emph {et~al.}(2018)\citenamefont
  {Bart\'ok}, \citenamefont {Kermode}, \citenamefont {Bernstein},\ and\
  \citenamefont {Cs\'anyi}}]{Bartok-PRX-2018}%
  \BibitemOpen
  \bibfield  {author} {\bibinfo {author} {\bibfnamefont {A.~P.}\ \bibnamefont
  {Bart\'ok}}, \bibinfo {author} {\bibfnamefont {J.}~\bibnamefont {Kermode}},
  \bibinfo {author} {\bibfnamefont {N.}~\bibnamefont {Bernstein}}, \ and\
  \bibinfo {author} {\bibfnamefont {G.}~\bibnamefont {Cs\'anyi}},\ }\bibfield
  {title} {\enquote {\bibinfo {title} {Machine learning a general-purpose
  interatomic potential for silicon},}\ }\href {\doibase
  10.1103/PhysRevX.8.041048} {\bibfield  {journal} {\bibinfo  {journal} {Phys.
  Rev. X}\ }\textbf {\bibinfo {volume} {8}},\ \bibinfo {pages} {041048}
  (\bibinfo {year} {2018})}\BibitemShut {NoStop}%
\bibitem [{\citenamefont {Herr}\ \emph {et~al.}(2018)\citenamefont {Herr},
  \citenamefont {Yao}, \citenamefont {McIntyre}, \citenamefont {Toth},\ and\
  \citenamefont {Parkhill}}]{herr2018metadynamics}%
  \BibitemOpen
  \bibfield  {author} {\bibinfo {author} {\bibfnamefont {J.~E.}\ \bibnamefont
  {Herr}}, \bibinfo {author} {\bibfnamefont {K.}~\bibnamefont {Yao}}, \bibinfo
  {author} {\bibfnamefont {R.}~\bibnamefont {McIntyre}}, \bibinfo {author}
  {\bibfnamefont {D.~W.}\ \bibnamefont {Toth}}, \ and\ \bibinfo {author}
  {\bibfnamefont {J.}~\bibnamefont {Parkhill}},\ }\bibfield  {title} {\enquote
  {\bibinfo {title} {Metadynamics for training neural network model
  chemistries: A competitive assessment},}\ }\href {\doibase
  https://doi.org/10.1063/1.5020067} {\bibfield  {journal} {\bibinfo  {journal}
  {J. Chem. Phys.}\ }\textbf {\bibinfo {volume} {148}},\ \bibinfo {pages}
  {241710} (\bibinfo {year} {2018})}\BibitemShut {NoStop}%
\bibitem [{\citenamefont {Huang}\ \emph {et~al.}(2018)\citenamefont {Huang},
  \citenamefont {Shang}, \citenamefont {Kang},\ and\ \citenamefont
  {Liu}}]{huang2018atomic}%
  \BibitemOpen
  \bibfield  {author} {\bibinfo {author} {\bibfnamefont {S.-D.}\ \bibnamefont
  {Huang}}, \bibinfo {author} {\bibfnamefont {C.}~\bibnamefont {Shang}},
  \bibinfo {author} {\bibfnamefont {P.-L.}\ \bibnamefont {Kang}}, \ and\
  \bibinfo {author} {\bibfnamefont {Z.-P.}\ \bibnamefont {Liu}},\ }\bibfield
  {title} {\enquote {\bibinfo {title} {Atomic structure of boron resolved using
  machine learning and global sampling},}\ }\href {\doibase
  https://doi.org/10.1039/C8SC03427C} {\bibfield  {journal} {\bibinfo
  {journal} {Chem. Sci.}\ }\textbf {\bibinfo {volume} {9}},\ \bibinfo {pages}
  {8644--8655} (\bibinfo {year} {2018})}\BibitemShut {NoStop}%
\bibitem [{\citenamefont {Kolsbjerg}, \citenamefont {Peterson},\ and\
  \citenamefont {Hammer}(2018)}]{Kolsbjerg-2018-PRB}%
  \BibitemOpen
  \bibfield  {author} {\bibinfo {author} {\bibfnamefont {E.~L.}\ \bibnamefont
  {Kolsbjerg}}, \bibinfo {author} {\bibfnamefont {A.~A.}\ \bibnamefont
  {Peterson}}, \ and\ \bibinfo {author} {\bibfnamefont {B.}~\bibnamefont
  {Hammer}},\ }\bibfield  {title} {\enquote {\bibinfo {title}
  {Neural-network-enhanced evolutionary algorithm applied to supported metal
  nanoparticles},}\ }\href {\doibase 10.1103/PhysRevB.97.195424} {\bibfield
  {journal} {\bibinfo  {journal} {Phys. Rev. B}\ }\textbf {\bibinfo {volume}
  {97}},\ \bibinfo {pages} {195424} (\bibinfo {year} {2018})}\BibitemShut
  {NoStop}%
\bibitem [{\citenamefont {Bart\'ok}, \citenamefont {Kondor},\ and\
  \citenamefont {Cs\'anyi}(2013)}]{Bartok-PRB-2013}%
  \BibitemOpen
  \bibfield  {author} {\bibinfo {author} {\bibfnamefont {A.~P.}\ \bibnamefont
  {Bart\'ok}}, \bibinfo {author} {\bibfnamefont {R.}~\bibnamefont {Kondor}}, \
  and\ \bibinfo {author} {\bibfnamefont {G.}~\bibnamefont {Cs\'anyi}},\
  }\bibfield  {title} {\enquote {\bibinfo {title} {On representing chemical
  environments},}\ }\href {\doibase 10.1103/PhysRevB.87.184115} {\bibfield
  {journal} {\bibinfo  {journal} {Phys. Rev. B}\ }\textbf {\bibinfo {volume}
  {87}},\ \bibinfo {pages} {184115} (\bibinfo {year} {2013})}\BibitemShut
  {NoStop}%
\bibitem [{\citenamefont {Li}, \citenamefont {Kermode},\ and\ \citenamefont
  {De~Vita}(2015)}]{Li-PRL-2015}%
  \BibitemOpen
  \bibfield  {author} {\bibinfo {author} {\bibfnamefont {Z.}~\bibnamefont
  {Li}}, \bibinfo {author} {\bibfnamefont {J.~R.}\ \bibnamefont {Kermode}}, \
  and\ \bibinfo {author} {\bibfnamefont {A.}~\bibnamefont {De~Vita}},\
  }\bibfield  {title} {\enquote {\bibinfo {title} {Molecular dynamics with
  on-the-fly machine learning of quantum-mechanical forces},}\ }\href {\doibase
  10.1103/PhysRevLett.114.096405} {\bibfield  {journal} {\bibinfo  {journal}
  {Phys. Rev. Lett.}\ }\textbf {\bibinfo {volume} {114}},\ \bibinfo {pages}
  {096405} (\bibinfo {year} {2015})}\BibitemShut {NoStop}%
\bibitem [{\citenamefont {Babaei}\ \emph {et~al.}(2019)\citenamefont {Babaei},
  \citenamefont {Guo}, \citenamefont {Hashemi},\ and\ \citenamefont
  {Lee}}]{BabaeiSi2019}%
  \BibitemOpen
  \bibfield  {author} {\bibinfo {author} {\bibfnamefont {H.}~\bibnamefont
  {Babaei}}, \bibinfo {author} {\bibfnamefont {R.}~\bibnamefont {Guo}},
  \bibinfo {author} {\bibfnamefont {A.}~\bibnamefont {Hashemi}}, \ and\
  \bibinfo {author} {\bibfnamefont {S.}~\bibnamefont {Lee}},\ }\bibfield
  {title} {\enquote {\bibinfo {title} {Machine-learning-based interatomic
  potential for phonon transport in perfect crystalline si and crystalline si
  with vacancies},}\ }\href {\doibase 10.1103/PhysRevMaterials.3.074603}
  {\bibfield  {journal} {\bibinfo  {journal} {Phys. Rev. Materials}\ }\textbf
  {\bibinfo {volume} {3}},\ \bibinfo {pages} {074603} (\bibinfo {year}
  {2019})}\BibitemShut {NoStop}%
\bibitem [{\citenamefont {Bonati}\ and\ \citenamefont
  {Parrinello}(2018)}]{Bonati-PRL-2018}%
  \BibitemOpen
  \bibfield  {author} {\bibinfo {author} {\bibfnamefont {L.}~\bibnamefont
  {Bonati}}\ and\ \bibinfo {author} {\bibfnamefont {M.}~\bibnamefont
  {Parrinello}},\ }\bibfield  {title} {\enquote {\bibinfo {title} {Silicon
  liquid structure and crystal nucleation from ab initio deep metadynamics},}\
  }\href {\doibase 10.1103/PhysRevLett.121.265701} {\bibfield  {journal}
  {\bibinfo  {journal} {Phys. Rev. Lett.}\ }\textbf {\bibinfo {volume} {121}},\
  \bibinfo {pages} {265701} (\bibinfo {year} {2018})}\BibitemShut {NoStop}%
\bibitem [{\citenamefont {Fredericks}, \citenamefont {Sayre},\ and\
  \citenamefont {Zhu}(2019)}]{pyxtal}%
  \BibitemOpen
  \bibfield  {author} {\bibinfo {author} {\bibfnamefont {S.}~\bibnamefont
  {Fredericks}}, \bibinfo {author} {\bibfnamefont {D.}~\bibnamefont {Sayre}}, \
  and\ \bibinfo {author} {\bibfnamefont {Q.}~\bibnamefont {Zhu}},\ }\bibfield
  {title} {\enquote {\bibinfo {title} {Pyxtal: a python library for crystal
  structure generation and symmetry analysis},}\ }\href@noop {} {\  (\bibinfo
  {year} {2019})},\ \Eprint {http://arxiv.org/abs/1911.11123} {arXiv:1911.11123
  [cond-mat.mtrl-sci]} \BibitemShut {NoStop}%
\bibitem [{\citenamefont {Larsen}\ \emph {et~al.}(2017)\citenamefont {Larsen},
  \citenamefont {Mortensen}, \citenamefont {Blomqvist}, \citenamefont
  {Castelli}, \citenamefont {Christensen}, \citenamefont {Du{\l}ak},
  \citenamefont {Friis}, \citenamefont {Groves}, \citenamefont {Hammer},
  \citenamefont {Hargus} \emph {et~al.}}]{larsen2017atomic}%
  \BibitemOpen
  \bibfield  {author} {\bibinfo {author} {\bibfnamefont {A.~H.}\ \bibnamefont
  {Larsen}}, \bibinfo {author} {\bibfnamefont {J.~J.}\ \bibnamefont
  {Mortensen}}, \bibinfo {author} {\bibfnamefont {J.}~\bibnamefont
  {Blomqvist}}, \bibinfo {author} {\bibfnamefont {I.~E.}\ \bibnamefont
  {Castelli}}, \bibinfo {author} {\bibfnamefont {R.}~\bibnamefont
  {Christensen}}, \bibinfo {author} {\bibfnamefont {M.}~\bibnamefont
  {Du{\l}ak}}, \bibinfo {author} {\bibfnamefont {J.}~\bibnamefont {Friis}},
  \bibinfo {author} {\bibfnamefont {M.~N.}\ \bibnamefont {Groves}}, \bibinfo
  {author} {\bibfnamefont {B.}~\bibnamefont {Hammer}}, \bibinfo {author}
  {\bibfnamefont {C.}~\bibnamefont {Hargus}},  \emph {et~al.},\ }\bibfield
  {title} {\enquote {\bibinfo {title} {The atomic simulation environment—a
  python library for working with atoms},}\ }\href {\doibase
  10.1088/1361-648X/aa680e} {\bibfield  {journal} {\bibinfo  {journal} {J.
  Phys. Condensed Matter}\ }\textbf {\bibinfo {volume} {29}},\ \bibinfo {pages}
  {273002} (\bibinfo {year} {2017})}\BibitemShut {NoStop}%
\bibitem [{\citenamefont {Kresse}\ and\ \citenamefont
  {Furthm{\"u}ller}(1996)}]{kresse1996efficient}%
  \BibitemOpen
  \bibfield  {author} {\bibinfo {author} {\bibfnamefont {G.}~\bibnamefont
  {Kresse}}\ and\ \bibinfo {author} {\bibfnamefont {J.}~\bibnamefont
  {Furthm{\"u}ller}},\ }\bibfield  {title} {\enquote {\bibinfo {title}
  {Efficient iterative schemes for ab initio total-energy calculations using a
  plane-wave basis set},}\ }\href {\doibase 10.1103/PhysRevB.54.11169}
  {\bibfield  {journal} {\bibinfo  {journal} {Phys. Rev. B}\ }\textbf {\bibinfo
  {volume} {54}},\ \bibinfo {pages} {11169} (\bibinfo {year}
  {1996})}\BibitemShut {NoStop}%
\bibitem [{\citenamefont {Bl{\"o}chl}(1994)}]{blochl1994projector}%
  \BibitemOpen
  \bibfield  {author} {\bibinfo {author} {\bibfnamefont {P.~E.}\ \bibnamefont
  {Bl{\"o}chl}},\ }\bibfield  {title} {\enquote {\bibinfo {title} {Projector
  augmented-wave method},}\ }\href {\doibase 10.1103/PhysRevB.50.17953}
  {\bibfield  {journal} {\bibinfo  {journal} {Phys. Rev. B}\ }\textbf {\bibinfo
  {volume} {50}},\ \bibinfo {pages} {17953} (\bibinfo {year}
  {1994})}\BibitemShut {NoStop}%
\bibitem [{\citenamefont {Perdew}, \citenamefont {Burke},\ and\ \citenamefont
  {Ernzerhof}(1996)}]{PBE-PRL-1996}%
  \BibitemOpen
  \bibfield  {author} {\bibinfo {author} {\bibfnamefont {J.~P.}\ \bibnamefont
  {Perdew}}, \bibinfo {author} {\bibfnamefont {K.}~\bibnamefont {Burke}}, \
  and\ \bibinfo {author} {\bibfnamefont {M.}~\bibnamefont {Ernzerhof}},\
  }\bibfield  {title} {\enquote {\bibinfo {title} {Generalized gradient
  approximation made simple},}\ }\href {\doibase 10.1103/PhysRevLett.77.3865}
  {\bibfield  {journal} {\bibinfo  {journal} {Phys. Rev. Lett.}\ }\textbf
  {\bibinfo {volume} {77}},\ \bibinfo {pages} {3865--3868} (\bibinfo {year}
  {1996})}\BibitemShut {NoStop}%
\bibitem [{\citenamefont {Gastegger}\ \emph {et~al.}(2018)\citenamefont
  {Gastegger}, \citenamefont {Schwiedrzik}, \citenamefont {Bittermann},
  \citenamefont {Berzsenyi},\ and\ \citenamefont
  {Marquetand}}]{Gastegger-2018-JCP}%
  \BibitemOpen
  \bibfield  {author} {\bibinfo {author} {\bibfnamefont {M.}~\bibnamefont
  {Gastegger}}, \bibinfo {author} {\bibfnamefont {L.}~\bibnamefont
  {Schwiedrzik}}, \bibinfo {author} {\bibfnamefont {M.}~\bibnamefont
  {Bittermann}}, \bibinfo {author} {\bibfnamefont {F.}~\bibnamefont
  {Berzsenyi}}, \ and\ \bibinfo {author} {\bibfnamefont {P.}~\bibnamefont
  {Marquetand}},\ }\bibfield  {title} {\enquote {\bibinfo {title}
  {wacsf—weighted atom-centered symmetry functions as descriptors in machine
  learning potentials},}\ }\href {\doibase 10.1063/1.5019667} {\bibfield
  {journal} {\bibinfo  {journal} {The Journal of chemical physics}\ }\textbf
  {\bibinfo {volume} {148}},\ \bibinfo {pages} {241709} (\bibinfo {year}
  {2018})}\BibitemShut {NoStop}%
\bibitem [{\citenamefont {Imbalzano}\ \emph {et~al.}(2018)\citenamefont
  {Imbalzano}, \citenamefont {Anelli}, \citenamefont {Giofre}, \citenamefont
  {Klees}, \citenamefont {Behler},\ and\ \citenamefont
  {Ceriotti}}]{Giulio-JCP-2018}%
  \BibitemOpen
  \bibfield  {author} {\bibinfo {author} {\bibfnamefont {G.}~\bibnamefont
  {Imbalzano}}, \bibinfo {author} {\bibfnamefont {A.}~\bibnamefont {Anelli}},
  \bibinfo {author} {\bibfnamefont {D.}~\bibnamefont {Giofre}}, \bibinfo
  {author} {\bibfnamefont {S.}~\bibnamefont {Klees}}, \bibinfo {author}
  {\bibfnamefont {J.}~\bibnamefont {Behler}}, \ and\ \bibinfo {author}
  {\bibfnamefont {M.}~\bibnamefont {Ceriotti}},\ }\bibfield  {title} {\enquote
  {\bibinfo {title} {Automatic selection of atomic fingerprints and reference
  configurations for machine-learning potentials},}\ }\href {\doibase
  10.1063/1.5024611} {\bibfield  {journal} {\bibinfo  {journal} {J. Chem.
  Phys.}\ }\textbf {\bibinfo {volume} {148}},\ \bibinfo {pages} {241730}
  (\bibinfo {year} {2018})}\BibitemShut {NoStop}%
\bibitem [{\citenamefont {Huan}\ \emph {et~al.}(2017)\citenamefont {Huan},
  \citenamefont {Batra}, \citenamefont {Chapman}, \citenamefont {Krishnan},
  \citenamefont {Chen},\ and\ \citenamefont {Ramprasad}}]{Huan-2017-NCM}%
  \BibitemOpen
  \bibfield  {author} {\bibinfo {author} {\bibfnamefont {T.~D.}\ \bibnamefont
  {Huan}}, \bibinfo {author} {\bibfnamefont {R.}~\bibnamefont {Batra}},
  \bibinfo {author} {\bibfnamefont {J.}~\bibnamefont {Chapman}}, \bibinfo
  {author} {\bibfnamefont {S.}~\bibnamefont {Krishnan}}, \bibinfo {author}
  {\bibfnamefont {L.}~\bibnamefont {Chen}}, \ and\ \bibinfo {author}
  {\bibfnamefont {R.}~\bibnamefont {Ramprasad}},\ }\bibfield  {title} {\enquote
  {\bibinfo {title} {A universal strategy for the creation of machine
  learning-based atomistic force fields},}\ }\href {\doibase
  10.1038/s41524-017-0042-y} {\bibfield  {journal} {\bibinfo  {journal} {npj
  Comput. Mater.}\ }\textbf {\bibinfo {volume} {3}} (\bibinfo {year} {2017}),\
  10.1038/s41524-017-0042-y}\BibitemShut {NoStop}%
\bibitem [{\citenamefont {Gao}, \citenamefont {Wang},\ and\ \citenamefont
  {Sun}(2019)}]{Gao-JCP-2019}%
  \BibitemOpen
  \bibfield  {author} {\bibinfo {author} {\bibfnamefont {H.}~\bibnamefont
  {Gao}}, \bibinfo {author} {\bibfnamefont {J.}~\bibnamefont {Wang}}, \ and\
  \bibinfo {author} {\bibfnamefont {J.}~\bibnamefont {Sun}},\ }\bibfield
  {title} {\enquote {\bibinfo {title} {Improve the performance of
  machine-learning potentials by optimizing descriptors},}\ }\href {\doibase
  10.1063/1.5097293} {\bibfield  {journal} {\bibinfo  {journal} {J. Chem.
  Phys.}\ }\textbf {\bibinfo {volume} {150}},\ \bibinfo {pages} {244110}
  (\bibinfo {year} {2019})}\BibitemShut {NoStop}%
\bibitem [{\citenamefont {Plimpton}(1995)}]{LAMMPS}%
  \BibitemOpen
  \bibfield  {author} {\bibinfo {author} {\bibfnamefont {S.}~\bibnamefont
  {Plimpton}},\ }\bibfield  {title} {\enquote {\bibinfo {title} {Fast parallel
  algorithms for short-range molecular dynamics},}\ }\href {\doibase
  https://doi.org/10.1006/jcph.1995.1039} {\bibfield  {journal} {\bibinfo
  {journal} {J. Comp. Phys.}\ }\textbf {\bibinfo {volume} {117}},\ \bibinfo
  {pages} {1 -- 19} (\bibinfo {year} {1995})}\BibitemShut {NoStop}%
\bibitem [{\citenamefont {Boyle}(2013)}]{PhysRevD.87.104006}%
  \BibitemOpen
  \bibfield  {author} {\bibinfo {author} {\bibfnamefont {M.}~\bibnamefont
  {Boyle}},\ }\bibfield  {title} {\enquote {\bibinfo {title} {Angular velocity
  of gravitational radiation from precessing binaries and the corotating
  frame},}\ }\href {\doibase 10.1103/PhysRevD.87.104006} {\bibfield  {journal}
  {\bibinfo  {journal} {Phys. Rev. D}\ }\textbf {\bibinfo {volume} {87}},\
  \bibinfo {pages} {104006} (\bibinfo {year} {2013})}\BibitemShut {NoStop}%
\bibitem [{\citenamefont {Oliphant}(2006)}]{oliphant2006guide}%
  \BibitemOpen
  \bibfield  {author} {\bibinfo {author} {\bibfnamefont {T.~E.}\ \bibnamefont
  {Oliphant}},\ }\href {http://www.numpy.org/} {\emph {\bibinfo {title} {A
  guide to NumPy}}},\ Vol.~\bibinfo {volume} {1}\ (\bibinfo  {publisher}
  {Trelgol Publishing USA},\ \bibinfo {year} {2006})\BibitemShut {NoStop}%
\bibitem [{\citenamefont {Kingma}\ and\ \citenamefont {Ba}(2014)}]{Adam}%
  \BibitemOpen
  \bibfield  {author} {\bibinfo {author} {\bibfnamefont {D.~P.}\ \bibnamefont
  {Kingma}}\ and\ \bibinfo {author} {\bibfnamefont {J.}~\bibnamefont {Ba}},\
  }\href {http://arxiv.org/abs/1412.6980} {\enquote {\bibinfo {title} {Adam: A
  method for stochastic optimization},}\ } (\bibinfo {year} {2014}),\ \bibinfo
  {note} {published as a conference paper at the 3rd International Conference
  for Learning Representations, San Diego, 2015}\BibitemShut {NoStop}%
\bibitem [{\citenamefont {{Virtanen}}\ \emph {et~al.}(2019)\citenamefont
  {{Virtanen}}, \citenamefont {{Gommers}}, \citenamefont {{Oliphant}},
  \citenamefont {{Haberland}}, \citenamefont {{Reddy}}, \citenamefont
  {{Cournapeau}}, \citenamefont {{Burovski}}, \citenamefont {{Peterson}},
  \citenamefont {{Weckesser}}, \citenamefont {{Bright}}, \citenamefont {{van
  der Walt}}, \citenamefont {{Brett}}, \citenamefont {{Wilson}}, \citenamefont
  {{Jarrod Millman}}, \citenamefont {{Mayorov}}, \citenamefont {{Nelson}},
  \citenamefont {{Jones}}, \citenamefont {{Kern}}, \citenamefont {{Larson}},
  \citenamefont {{Carey}}, \citenamefont {{Polat}}, \citenamefont {{Feng}},
  \citenamefont {{Moore}}, \citenamefont {{Vand erPlas}}, \citenamefont
  {{Laxalde}}, \citenamefont {{Perktold}}, \citenamefont {{Cimrman}},
  \citenamefont {{Henriksen}}, \citenamefont {{Quintero}}, \citenamefont
  {{Harris}}, \citenamefont {{Archibald}}, \citenamefont {{Ribeiro}},
  \citenamefont {{Pedregosa}}, \citenamefont {{van Mulbregt}},\ and\
  \citenamefont {{Contributors}}}]{scipy}%
  \BibitemOpen
  \bibfield  {author} {\bibinfo {author} {\bibfnamefont {P.}~\bibnamefont
  {{Virtanen}}}, \bibinfo {author} {\bibfnamefont {R.}~\bibnamefont
  {{Gommers}}}, \bibinfo {author} {\bibfnamefont {T.~E.}\ \bibnamefont
  {{Oliphant}}}, \bibinfo {author} {\bibfnamefont {M.}~\bibnamefont
  {{Haberland}}}, \bibinfo {author} {\bibfnamefont {T.}~\bibnamefont
  {{Reddy}}}, \bibinfo {author} {\bibfnamefont {D.}~\bibnamefont
  {{Cournapeau}}}, \bibinfo {author} {\bibfnamefont {E.}~\bibnamefont
  {{Burovski}}}, \bibinfo {author} {\bibfnamefont {P.}~\bibnamefont
  {{Peterson}}}, \bibinfo {author} {\bibfnamefont {W.}~\bibnamefont
  {{Weckesser}}}, \bibinfo {author} {\bibfnamefont {J.}~\bibnamefont
  {{Bright}}}, \bibinfo {author} {\bibfnamefont {S.~J.}\ \bibnamefont {{van der
  Walt}}}, \bibinfo {author} {\bibfnamefont {M.}~\bibnamefont {{Brett}}},
  \bibinfo {author} {\bibfnamefont {J.}~\bibnamefont {{Wilson}}}, \bibinfo
  {author} {\bibfnamefont {K.}~\bibnamefont {{Jarrod Millman}}}, \bibinfo
  {author} {\bibfnamefont {N.}~\bibnamefont {{Mayorov}}}, \bibinfo {author}
  {\bibfnamefont {A.~R.~J.}\ \bibnamefont {{Nelson}}}, \bibinfo {author}
  {\bibfnamefont {E.}~\bibnamefont {{Jones}}}, \bibinfo {author} {\bibfnamefont
  {R.}~\bibnamefont {{Kern}}}, \bibinfo {author} {\bibfnamefont
  {E.}~\bibnamefont {{Larson}}}, \bibinfo {author} {\bibfnamefont
  {C.}~\bibnamefont {{Carey}}}, \bibinfo {author} {\bibfnamefont
  {{\.I}.}~\bibnamefont {{Polat}}}, \bibinfo {author} {\bibfnamefont
  {Y.}~\bibnamefont {{Feng}}}, \bibinfo {author} {\bibfnamefont {E.~W.}\
  \bibnamefont {{Moore}}}, \bibinfo {author} {\bibfnamefont {J.}~\bibnamefont
  {{Vand erPlas}}}, \bibinfo {author} {\bibfnamefont {D.}~\bibnamefont
  {{Laxalde}}}, \bibinfo {author} {\bibfnamefont {J.}~\bibnamefont
  {{Perktold}}}, \bibinfo {author} {\bibfnamefont {R.}~\bibnamefont
  {{Cimrman}}}, \bibinfo {author} {\bibfnamefont {I.}~\bibnamefont
  {{Henriksen}}}, \bibinfo {author} {\bibfnamefont {E.~A.}\ \bibnamefont
  {{Quintero}}}, \bibinfo {author} {\bibfnamefont {C.~R.}\ \bibnamefont
  {{Harris}}}, \bibinfo {author} {\bibfnamefont {A.~M.}\ \bibnamefont
  {{Archibald}}}, \bibinfo {author} {\bibfnamefont {A.~H.}\ \bibnamefont
  {{Ribeiro}}}, \bibinfo {author} {\bibfnamefont {F.}~\bibnamefont
  {{Pedregosa}}}, \bibinfo {author} {\bibfnamefont {P.}~\bibnamefont {{van
  Mulbregt}}}, \ and\ \bibinfo {author} {\bibfnamefont {S.~.~.}\ \bibnamefont
  {{Contributors}}},\ }\bibfield  {title} {\enquote {\bibinfo {title} {{SciPy
  1.0--Fundamental Algorithms for Scientific Computing in Python}},}\ }\href
  {https://arxiv.org/abs/1907.10121} {\bibfield  {journal} {\bibinfo  {journal}
  {arXiv e-prints}\ ,\ \bibinfo {pages} {arXiv:1907.10121}} (\bibinfo {year}
  {2019})},\ \Eprint {http://arxiv.org/abs/1907.10121} {arXiv:1907.10121
  [cs.MS]} \BibitemShut {NoStop}%
\bibitem [{\citenamefont {Zhu}\ \emph {et~al.}(1997)\citenamefont {Zhu},
  \citenamefont {Byrd}, \citenamefont {Lu},\ and\ \citenamefont
  {Nocedal}}]{L-BFGS}%
  \BibitemOpen
  \bibfield  {author} {\bibinfo {author} {\bibfnamefont {C.}~\bibnamefont
  {Zhu}}, \bibinfo {author} {\bibfnamefont {R.~H.}\ \bibnamefont {Byrd}},
  \bibinfo {author} {\bibfnamefont {P.}~\bibnamefont {Lu}}, \ and\ \bibinfo
  {author} {\bibfnamefont {J.}~\bibnamefont {Nocedal}},\ }\bibfield  {title}
  {\enquote {\bibinfo {title} {Algorithm 778: L-bfgs-b: Fortran subroutines for
  large-scale bound-constrained optimization},}\ }\href {\doibase
  10.1145/279232.279236} {\bibfield  {journal} {\bibinfo  {journal} {ACM Trans.
  Math. Softw.}\ }\textbf {\bibinfo {volume} {23}},\ \bibinfo {pages}
  {550--560} (\bibinfo {year} {1997})}\BibitemShut {NoStop}%
\bibitem [{\citenamefont {Kuritz}, \citenamefont {Gordon},\ and\ \citenamefont
  {Natan}(2018)}]{kuritz2018size}%
  \BibitemOpen
  \bibfield  {author} {\bibinfo {author} {\bibfnamefont {N.}~\bibnamefont
  {Kuritz}}, \bibinfo {author} {\bibfnamefont {G.}~\bibnamefont {Gordon}}, \
  and\ \bibinfo {author} {\bibfnamefont {A.}~\bibnamefont {Natan}},\ }\bibfield
   {title} {\enquote {\bibinfo {title} {Size and temperature transferability of
  direct and local deep neural networks for atomic forces},}\ }\href {\doibase
  10.1103/PhysRevB.98.094109} {\bibfield  {journal} {\bibinfo  {journal}
  {Physical Review B}\ }\textbf {\bibinfo {volume} {98}},\ \bibinfo {pages}
  {094109} (\bibinfo {year} {2018})}\BibitemShut {NoStop}%
\bibitem [{\citenamefont {Schall}, \citenamefont {Gao},\ and\ \citenamefont
  {Harrison}(2008)}]{schall2008elastic}%
  \BibitemOpen
  \bibfield  {author} {\bibinfo {author} {\bibfnamefont {J.~D.}\ \bibnamefont
  {Schall}}, \bibinfo {author} {\bibfnamefont {G.}~\bibnamefont {Gao}}, \ and\
  \bibinfo {author} {\bibfnamefont {J.~A.}\ \bibnamefont {Harrison}},\
  }\bibfield  {title} {\enquote {\bibinfo {title} {Elastic constants of silicon
  materials calculated as a function of temperature using a parametrization of
  the second-generation reactive empirical bond-order potential},}\ }\href
  {\doibase 10.1103/PhysRevB.77.115209} {\bibfield  {journal} {\bibinfo
  {journal} {Phys. Rev. B}\ }\textbf {\bibinfo {volume} {77}},\ \bibinfo
  {pages} {115209} (\bibinfo {year} {2008})}\BibitemShut {NoStop}%
\bibitem [{\citenamefont {Lyakhov}\ \emph {et~al.}(2013)\citenamefont
  {Lyakhov}, \citenamefont {Oganov}, \citenamefont {Stokes},\ and\
  \citenamefont {Zhu}}]{Lyakhov-CPC-2013}%
  \BibitemOpen
  \bibfield  {author} {\bibinfo {author} {\bibfnamefont {A.~O.}\ \bibnamefont
  {Lyakhov}}, \bibinfo {author} {\bibfnamefont {A.~R.}\ \bibnamefont {Oganov}},
  \bibinfo {author} {\bibfnamefont {H.~T.}\ \bibnamefont {Stokes}}, \ and\
  \bibinfo {author} {\bibfnamefont {Q.}~\bibnamefont {Zhu}},\ }\bibfield
  {title} {\enquote {\bibinfo {title} {New developments in evolutionary
  structure prediction algorithm uspex},}\ }\href {\doibase
  10.1016/j.cpc.2012.12.009} {\bibfield  {journal} {\bibinfo  {journal}
  {Comput. Phys. Commun.}\ }\textbf {\bibinfo {volume} {184}},\ \bibinfo
  {pages} {1172--1182} (\bibinfo {year} {2013})}\BibitemShut {NoStop}%
\bibitem [{\citenamefont {Huang}\ \emph {et~al.}(2019)\citenamefont {Huang},
  \citenamefont {Kang}, \citenamefont {Goddard},\ and\ \citenamefont
  {Wang}}]{Huang-PRB-2019}%
  \BibitemOpen
  \bibfield  {author} {\bibinfo {author} {\bibfnamefont {Y.}~\bibnamefont
  {Huang}}, \bibinfo {author} {\bibfnamefont {J.}~\bibnamefont {Kang}},
  \bibinfo {author} {\bibfnamefont {W.~A.}\ \bibnamefont {Goddard}}, \ and\
  \bibinfo {author} {\bibfnamefont {L.-W.}\ \bibnamefont {Wang}},\ }\bibfield
  {title} {\enquote {\bibinfo {title} {Density functional theory based neural
  network force fields from energy decompositions},}\ }\href {\doibase
  10.1103/PhysRevB.99.064103} {\bibfield  {journal} {\bibinfo  {journal} {Phys.
  Rev. B}\ }\textbf {\bibinfo {volume} {99}},\ \bibinfo {pages} {064103}
  (\bibinfo {year} {2019})}\BibitemShut {NoStop}%
\bibitem [{\citenamefont {Yoo}\ \emph {et~al.}(2019)\citenamefont {Yoo},
  \citenamefont {Lee}, \citenamefont {Jeong}, \citenamefont {Lee},
  \citenamefont {Watanabe},\ and\ \citenamefont {Han}}]{Lee-PRM-2019}%
  \BibitemOpen
  \bibfield  {author} {\bibinfo {author} {\bibfnamefont {D.}~\bibnamefont
  {Yoo}}, \bibinfo {author} {\bibfnamefont {K.}~\bibnamefont {Lee}}, \bibinfo
  {author} {\bibfnamefont {W.}~\bibnamefont {Jeong}}, \bibinfo {author}
  {\bibfnamefont {D.}~\bibnamefont {Lee}}, \bibinfo {author} {\bibfnamefont
  {S.}~\bibnamefont {Watanabe}}, \ and\ \bibinfo {author} {\bibfnamefont
  {S.}~\bibnamefont {Han}},\ }\bibfield  {title} {\enquote {\bibinfo {title}
  {Atomic energy mapping of neural network potential},}\ }\href {\doibase
  10.1103/PhysRevMaterials.3.093802} {\bibfield  {journal} {\bibinfo  {journal}
  {Phys. Rev. Mater.}\ }\textbf {\bibinfo {volume} {3}},\ \bibinfo {pages}
  {093802} (\bibinfo {year} {2019})}\BibitemShut {NoStop}%
\bibitem [{\citenamefont {Deringer}\ and\ \citenamefont
  {Cs\'anyi}(2017)}]{Deringer-carbon-PRB}%
  \BibitemOpen
  \bibfield  {author} {\bibinfo {author} {\bibfnamefont {V.~L.}\ \bibnamefont
  {Deringer}}\ and\ \bibinfo {author} {\bibfnamefont {G.}~\bibnamefont
  {Cs\'anyi}},\ }\bibfield  {title} {\enquote {\bibinfo {title} {Machine
  learning based interatomic potential for amorphous carbon},}\ }\href
  {\doibase 10.1103/PhysRevB.95.094203} {\bibfield  {journal} {\bibinfo
  {journal} {Phys. Rev. B}\ }\textbf {\bibinfo {volume} {95}},\ \bibinfo
  {pages} {094203} (\bibinfo {year} {2017})}\BibitemShut {NoStop}%
\end{thebibliography}%

\end{document}